\documentclass[journal=jpclcd,manuscript=letter]{achemso}

\usepackage{amsmath}
\usepackage{bm}
\usepackage{graphicx}
\usepackage[version=4]{mhchem}

\newcommand{\dd}{\mathop{}\!\text{d}}

\author{Zehua Chen}
\author{Yang Yang}
\affiliation{Theoretical Chemistry Institute and Department of Chemistry,
University of Wisconsin-Madison,
1101 University Avenue, Madison, Wisconsin 53706, United States}
\email{yyang222@wisc.edu}

\title{Incorporating Nuclear Quantum Effects in Molecular Dynamics
with a Constrained Minimized Energy Surface}


\begin{document}

\begin{tocentry}
  \includegraphics[width=\linewidth]{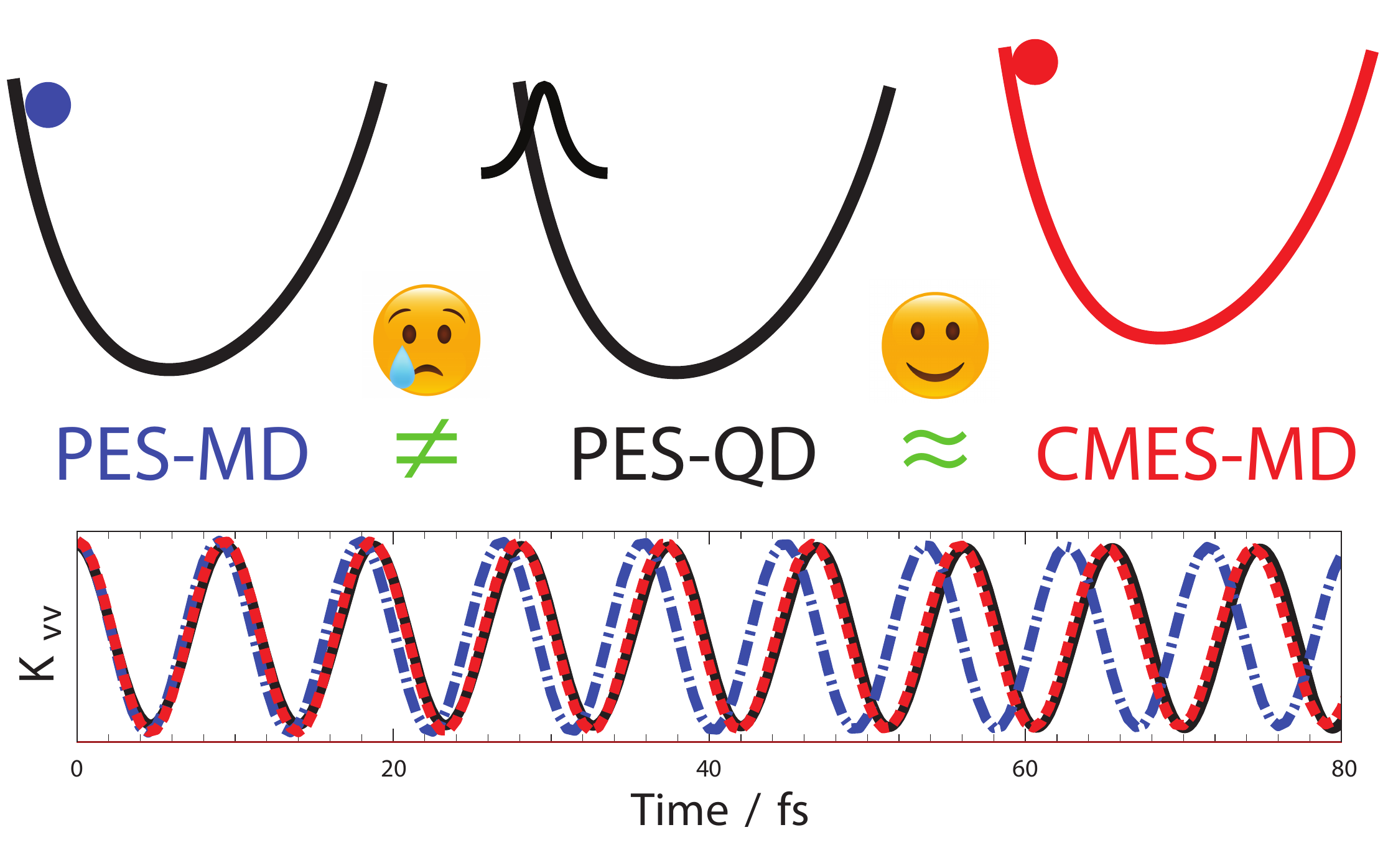}
\end{tocentry}

\begin{abstract}
  The accurate incorporation of nuclear quantum effects in large-scale
  molecular dynamics (MD) simulations remains a significant challenge.
  Recently, we combined constrained nuclear-electronic orbital (CNEO) theory
  with classical MD and obtained a new approach (CNEO-MD)
  that can accurately and efficiently incorporate nuclear quantum effects into
  classical simulations.
  In this Letter, we provide the theoretical foundation for CNEO-MD by developing
  an alternative formulation of the equations of motion for MD.
  In this new formulation, the expectation values of quantum nuclear positions
  evolve classically on an effective energy surface that is obtained from a
  constrained energy minimization procedure when solving for the quantum nuclear
  wave function, thus enabling the incorporation of nuclear quantum effects in
  classical MD simulations. For comparison with other existing
  approaches, we examined a series of model systems and found that this new MD
  approach is significantly more accurate than the conventional way of
  performing classical MD,
  and it also generally outperforms centroid MD and ring-polymer
  MD in describing vibrations in these model systems.
\end{abstract}

Nuclear quantum effects (NQEs)
\cite{Markland181}
have a great impact on the structural, thermodynamical, and kinetic properties
of a wide range of chemical and biological systems.
\cite{Pereyaslavets188878}
They usually include zero-point and tunneling effects and are significant when
light nuclei, such as hydrogen, are present. The accurate incorporation of NQEs
in molecular simulations is important for understanding many fundamental
properties but remains a significant challenge for large-scale molecular
simulations. For example, the anomalous properties of water are closely related
to the NQEs of the complex hydrogen bond network
\cite{Nilsson158998, Ceriotti167529}
and thus cannot be fully explained with conventional classical molecular dynamics (MD)
without an accurate inclusion of NQEs.
\cite{Zhang21236001}

There have been many theoretical developments in the incorporation of NQEs in
molecular simulations.
Quantum wave packet dynamics is based on the exact time evolution of a quantum
system according to the time-dependent Schr\"odinger equation and can give
theoretical predictions that accurately match experiments.
\cite{Meyer9073, Fleck76129, Weichman17950, Chen209202, Chen21936}
Quantum trajectory methods
\cite{Lopreore995190}
are based on the de Broglie-Bohm formulation of quantum mechanics,
\cite{de1926possibilite, de1927structure, Madelung27322, Bohm52166, Bohm52180, schleich2013schrodinger}
which attributes quantum effects to the quantum potential, and with a reasonable
approximation to the quantum potential, quantum trajectory methods have been
applied to many model systems and give accurate results.
\cite{Garashchuk03358, Garashchuk041181}
Multicomponent quantum theories
\cite{Thomas6990, Capitani82568, Tachikawa98437, Kreibich012984, Webb024106, Ishimoto092677, Pavosevic204222}
also include NQEs by simultaneously treating both electrons and key nuclei quantum mechanically,
which do not rely on conventional Born-Oppenheimer potential energy surfaces (PESs). Their
real-time dynamics simulations.
\cite{Abedi10123002, Suzuki1529271, Zhao204052, Zhao20224111, Tao214131}
can be performed through the
quantum time evolution of multicomponent wave functions or density matrices,
which have been used to study
practical chemical problems, such as proton transfer processes.
\cite{Zhao213497}

Although the aforementioned methods are highly accurate in describing NQEs,
they are often hindered by their high computational costs in large molecular
or bulk systems. This challenge has been partially addressed using
methods  based on classical simulations.
Some empirical force fields
\cite{Grigera018064}
have been used to include NQEs implicitly and have been able to treat hydrogens
and deuteriums differently in water.
\cite{BenAbu211}
The generalized Langevin equation thermostat with optimized parameters can also
include NQEs and has been applied to obtain several static properties.
\cite{Ceriotti09020601, Ceriotti09030603}
The semiclassical initial value representation (SC-IVR)
\cite{Wang989726}
and its approximate variants
\cite{Sun987064, Liu09074113}
can also produce accurate time correlation functions with classical simulations.
Methods based on the path integral formulation of quantum mechanics
\cite{Feynman10, feynman1972statistical}
are most popular for the inclusion of NQEs.
Via simultaneous simulation of a set of coupled replicas for a system, path
integral molecular dynamics (PIMD)
\cite{Berne98, tuckerman2002path}
can capture NQEs and accurately describe the static properties of the
system.
\cite{Herrero14233201}
Its extensions such as
centroid molecular dynamics (CMD)
\cite{Cao945106, Jang992371}
and
ring-polymer molecular dynamics (RPMD)
\cite{Craig043368}
can describe dynamical properties using approximate time correlation functions.
However, while dynamical properties from CMD and RPMD are considerably
more accurate than those from conventional MD, challenges still exist with
the curvature problem in CMD and spurious frequencies in RPMD.
Both of these problems can lead to unreliable vibrational spectra,
\cite{Witt09194510}
although several recent developments can mitigate them to some extent,
including thermostatted RPMD,
\cite{Rossi14234116, Rossi18102301}
Matsubara dynamics,
\cite{Hele15134103}
and quasicentroid molecular dynamics.
\cite{Haggard21174120,Fletcher21231101}
Furthermore, in contrast with PIMD, which has many techniques developed
to make it as efficient as conventional MD,
\cite{Kapil16054111, Marsalek16054112, Uhl16054101, Xue2110677}
currently there are only a limited number of techniques available
\cite{Gui225203}
to accelerate RPMD/CMD simulations other than massive parallelization,
and thus, the efficient simulation of dynamical properties remains a challenge.

In this Letter, we present an alternative formulation of the equations of
motion for classical molecular simulations, with which
NQEs can be described using an effective PES that is in practice
approximated by a constrained minimized energy surface (CMES).
This formulation serves as the theoretical foundation for our recently
developed MD approach based on constrained nuclear-electronic orbital theory
(CNEO-MD). For comparison with existing approaches, we examine a series of
model systems. We first show that CMES-MD remains exact for the harmonic
oscillator model. Then, with a Morse
oscillator model and a quartic double-well potential model, we show that
CMES-MD is generally much more accurate in describing vibrations and tunneling
effects than the conventional way of performing
classical MD and is comparable to or slightly better than CMD and RPMD.

We start with the polar representation of a time-dependent wave function
$\psi (\bm{x}, t) = A (\bm{x}, t) \exp (\mathrm{i} S (\bm{x}, t) / \hbar)$,
where amplitude part
$A$
and phase part
$S$
are both real. For the sake of simplicity, we assume in our derivation that
there is only one quantum particle, but this formulation can be
easily generalized to multiple quantum particle cases if the particles can be
assumed to be distinguishable, such as nuclei in regular molecular and bulk systems.
Additionally, we will assume that no magnetic field is present,
although we note that magnetic fields can be important on some occasions
and the corresponding formulation can be explored in the future.
With the polar representation, the kinetic energy can be decomposed into two
terms
\begin{align}
  \langle \hat{T} \rangle (t)
  & =   \int \dd \bm{x} A (\bm{x}, t)
  \frac{(- \mathrm{i} \hbar \nabla)^2}{2m} A (\bm{x}, t) \nonumber \\
  & \quad + \frac{1}{2m} \int \dd \bm{x} A^2 (\bm{x}, t)
  [\nabla S (\bm{x}, t)]^2 .
\end{align}
The first term is the kinetic energy evaluated with amplitude function
$A$
only. Because
$A$
is associated with the real space probability density distribution with
$\rho(\bm{x}, t) = A^2(\bm{x}, t)$,
this term can be perceived as the kinetic energy due to quantum delocalization,
or the zero-point kinetic energy. In the second term, the key quantity
$\nabla S$
is associated with the observable momentum and is related to the momentum field
in Bohmian mechanics
\cite{de1926possibilite, de1927structure, Madelung27322, Bohm52166, Bohm52180, schleich2013schrodinger}
with the definition
$\bm{p} (\bm{x}, t) = \nabla S (\bm{x}, t)$.
Because
$A^2(\bm{x}, t)$
is the probability density, this term can be viewed as
the kinetic energy associated with the observable momentum
$\bm{p}$.

We define the variance of the observable momentum as the variance of the
momentum field:
\begin{equation}
  \sigma_{\bm{p}}^2 (t) \equiv \int \dd \bm{x} A^2 (\bm{x}, t)
  [\nabla S (\bm{x}, t)]^2 - \langle \hat{\bm{p}} \rangle^2 (t) ,
  \label{eqn:sigma}
\end{equation}
then the kinetic energy can be further expressed as
\begin{equation}
  \langle \hat{T} \rangle (t) =
  \langle A (t) | \hat{T} | A (t) \rangle +
  \frac{\langle \hat{\bm{p}} \rangle^2 (t)}{2 m} +
  \frac{\sigma_{\bm{p}}^2 (t)}{2 m} .
  \label{eqn:kinetic_energy}
\end{equation}
The terms in eq~\ref{eqn:kinetic_energy} correspond to
the zero-point kinetic energy,
the classical kinetic energy associated with the expectation value of
the observable momentum,
and an energy contribution from the variance of the observable momentum field,
respectively.

Another way of expressing the kinetic energy is simply
$\langle \hat{T} \rangle (t) = \langle \hat{H} \rangle (t) - \langle \hat{V} \rangle (t)$,
which can be plugged into the left side of eq~\ref{eqn:kinetic_energy}.
Then by taking the time derivative on both sides of the equation, one can
simplify it to
\begin{equation}
  \frac{\langle \hat{\bm{p}} \rangle}{m} \cdot
  \frac{\dd \langle \hat{\bm{p}} \rangle}{\dd t}
  = \left\langle \frac{\partial V}{\partial t} \right\rangle -
  \frac{\dd}{\dd t} \langle A (t) | \hat{H} (t) | A (t) \rangle -
  \frac{\dd}{\dd t} \frac{\sigma_{\bm{p}}^2}{2 m}.
  \label{eqn:exact_eom}
\end{equation}
Note that we have used the relationship
$\dd \langle \hat{H} \rangle (t) / \dd t = \langle \partial V / \partial t \rangle$
in the derivation.
Equation~\ref{eqn:exact_eom} relates the time dependence of momentum to
the time dependence of energetic terms.
While eq~\ref{eqn:exact_eom} is exact,
to make it into an equation of motion that can be practically
used in MD simulations, we next proceed with an approximation that builds a
connection between quantum states and the classical phase space.

Conventionally, when assuming the potential is slowly varying in space,
we have the Ehrenfest theorem that provides
a connection between the classical Newtonian dynamics in phase space
$(\bm{X},\bm{P})$
and the evolution of quantum expectation values of position and momentum
$(\langle \hat{\bm{x}} \rangle, \langle \hat{\bm{p}} \rangle)$.
Here we build on the same mapping philosophy but instead of assuming the behavior of the potential,
we approximate the quantum state as the energy-minimized state for a given phase space point.
That is, when the system is at a particular phase space point given by an expectation
position and an expectation momentum, i.e.,
$(\langle \hat{\bm{x}} \rangle, \langle \hat{\bm{p}} \rangle) = (\bm{X},\bm{P})$,
quantum state
$|\psi\rangle$
always adapts to the energy-minimized state for that phase space point.
We note that this approximation
is an adiabatic approximation and is not trivially justifiable;
however, to keep the flow of the derivation, we leave discussions of its
applicability as well as limitations for the later part of this Letter.

Under this adiabatic approximation, quantum state $| \psi \rangle$
becomes an explicit function of
$(\bm{X},\bm{P})$
and an implicit function of time $t$, i.e.,
$| \psi \rangle (\bm{X}(t),\bm{P}(t))$.
At a particular phase space point
$(\bm{X}(t),\bm{P}(t))$
to which the system evolves at time $t$, the state can be obtained
with a constrained energy minimization procedure. The corresponding Lagrangian is
\begin{align}
  \mathcal{L} &= \langle \psi | \hat{H} (t) | \psi \rangle
  + \bm{f} \cdot (\langle \psi | \hat{\bm{x}} | \psi \rangle - \bm{X} (t)) \nonumber \\
  & \quad - \bm{v} \cdot (\langle \psi | \hat{\bm{p}} | \psi \rangle - \bm{P} (t))
  - \tilde{E}(\langle \psi | \psi \rangle - 1),
\end{align}
where
$\bm{f}$
is the Lagrange multiplier associated with the expectation position,
$\bm{v}$
is the Lagrange multiplier associated with the expectation momentum,
and
$\tilde{E}$
is the Lagrange multiplier associated with the wave function normalization.
This Lagrangian can be further expressed in terms of $A$ and $S$ by
\begin{align}
  \mathcal{L} & = \langle A | \hat{H} (t) | A \rangle
  + \frac{1}{2m} \int \dd \bm{x} A^2 (\nabla S)^2 \nonumber \\
  & \quad + \bm{f} \cdot (\langle A | \hat{\bm{x}} | A \rangle - \bm{X} (t))
  - \bm{v} \cdot \left(\int \dd \bm{x} A^2 \nabla S - \bm{P} (t)\right) \nonumber \\
  & \quad - \tilde{E}(\langle A | A \rangle - 1).
\end{align}
Making the Lagrangian function stationary with respect to the variation of
$\nabla S$ and $A$ leads to
$A^2(\nabla S/m - \bm{v})=0$
and
\begin{equation}
  \left[\hat{H} (t) + \frac{(\nabla S)^2}{2m} + \bm{f} \cdot \hat{\bm{x}} - \bm{v} \cdot \nabla S\right]
  | A \rangle = \tilde{E} | A \rangle.
\end{equation}
Further combining these equations with the expectation position constraint,
the expectation momentum constraint, and the normalization constraint gives
$\bm{v}=\bm{P}(t)/m$, $\nabla S(\bm{x},t) = m \bm{v} = \bm{P}(t)$,
and then the eigenvalue equation can be simplified to
\begin{equation}
  [\hat{H} (t) + \bm{f} \cdot \hat{\bm{x}}] | A \rangle =
  \left(\tilde{E} + \frac{\bm{P}^2}{2m}\right) | A \rangle.
  \label{eqn:constraint}
\end{equation}
The eigenvalue $\tilde{E} + \bm{P}^2/2m$, eigenstate $| A \rangle$,
and the Lagrange multiplier $\bm{f}$ can be solved under the expectation
position and normalization constraints for $| A \rangle$. Note that interestingly,
the solution of amplitude function $A$ depends only on the
expectation position constraint,
and the expectation momentum constraint affects only phase function $S$.

The fact that the constrained minimization requires $\nabla S$ to agree with the
momentum expectation value
($\nabla S(\bm{x},t) = \bm{P}(t)$) naturally leads to $\sigma_{\bm{p}}^2 = 0$
according to the definition in eq~\ref{eqn:sigma},
and with the quantum state
$|A\rangle$
obtained as an explicit function of
$\bm{X}$ and thus an implicit function of $t$,
we can simplify eq~\ref{eqn:exact_eom} into
\begin{align}
  \frac{\langle \hat{\bm{p}} \rangle}{m} \cdot
  \frac{\dd \langle \hat{\bm{p}} \rangle}{\dd t}
  & \approx -
  \left\langle \frac{\dd A}{\dd t} \middle| \hat{H} (t) \middle| A \right\rangle
  - \left\langle A
  \middle| \hat{H} (t) \middle| \frac{\dd A}{\dd t} \right\rangle \\
  & = -
  \frac{\dd \bm{X}}{\dd t} \cdot
  \Big[
    \langle \nabla_{\bm{X}} A | \hat{H} (t) | A \rangle \nonumber \\
  & \quad + \langle A | \hat{H} (t) | \nabla_{\bm{X}} A \rangle
  \Big] \\
  & = -
  \frac{\langle \hat{\bm{p}} \rangle}{m} \cdot
  \nabla_{\bm{X}} \langle A | \hat{H} (t) | A \rangle.
  \label{eqn:second_last}
\end{align}
Note that here we have used
$\dd \bm{X} / \dd t = \dd \langle\hat{\bm{x}}\rangle / \dd t =
\langle\hat{\bm{p}}\rangle / m$.
According to classical mechanics, it is natural to assume that the change in
$\langle\hat{\bm{p}}\rangle$
should have an opposite direction to the energy gradient term
($\nabla_{\bm{X}} \langle A | \hat{H} (t) | A \rangle$);
therefore, the common prefactor
$|\langle\hat{\bm{p}}\rangle|/m$
can be dropped, and we arrive at the final expression
\begin{equation}
  \frac{\dd \langle \hat{\bm{p}} \rangle}{\dd t}
  \approx
  - \nabla_{\bm{X}} \langle A | \hat{H} (t) | A \rangle
  \equiv
  - \nabla_{\bm{X}}
  V^{\text{CMES}}(\bm{X}),
  \label{eqn:eom}
\end{equation}
where
$V^{\text{CMES}}$
is the constrained minimized energy surface associated with
amplitude part $| A \rangle$.
It can also be viewed as an effective potential energy surface that includes
not only the potential energy but also the quantum delocalization kinetic energy.
Equation~\ref{eqn:eom},
together with
$\dd \langle\hat{\bm{x}}\rangle / \dd t = \langle\hat{\bm{p}}\rangle / m$,
forms the equations of motion for CMES-MD.
These equations of motion present an alternative way of performing MD simulations
but with NQEs incorporated.
They are highly similar in structure to Newton's equations used in conventional
MD simulations, with the difference that the time evolution is now on the
quantum expectation values of positions and momenta rather than the classical ones.

We note that there have been prior works
\cite{Ramirez983303,Ramirez993339,Ramirez994456}
that arrived at the same equations of motion within the framework of Feynman's
path-integral formulation of quantum mechanics. Within the path-integral framework,
the effective potential guides the motion of the ring-polymer centroid
and is claimed to be equal to the zero-temperature limit of the centroid potential for CMD.
\cite{Ramirez983303,Ramirez993339}
Therefore, it has been used to gain insight into the behavior of CMD.
In our work presented here, with a formal derivation from the
conventional formulation of quantum mechanics,
these equations provide a new way of performing classical molecular simulations
with the effective potential utilized to guide the classical motion of
the quantum expectation values.

Additionally, we note that
this formulation serves as the theoretical foundation for our recently published work
of CNEO-MD,
\cite{Xu224039}
in which constrained nuclear electronic orbital density functional theory (CNEO-DFT)
\cite{Xu20074106, Xu20084107, Xu21244110}
is employed to obtain CMESs for practical molecular systems and used for MD simulations.
CNEO-MD is a generalization of CMES-MD when the electronic part is also explicitly
considered in the energy minimization procedure. A more detailed derivation of their
connections can be found in section~S3 of the Supporting Information.
With a series of gas phase molecules,
we have demonstrated the excellent agreement between the CNEO-MD vibrational spectra and
the experimental spectra.
\cite{Xu224039}
Specifically for highly anharmonic \ce{O-H}
and \ce{C-H} stretching modes, CNEO-MD
significantly outperforms conventional DFT-based {\it ab initio} molecular dynamics,
with errors in peak positions reduced by 1 order of magnitude,
but at essentially the same computational cost.

For comparison with other existing approaches, herein,
we investigate the model systems of a harmonic oscillator,
a Morse oscillator and a quartic double-well potential.
These model systems are chosen because they have easily
accessible exact quantum solutions, avoid errors associated with real systems such
as electron correlations, and are affordable for CMD and RPMD simulations.
For these model systems,
we scan a set of discrete $f$ values in eq~\ref{eqn:constraint} to solve for the CMESs.
Specifically, for each $f$, the constrained eigenvalue equation is solved
numerically on a grid and the energy as a function of the corresponding nuclear expectation
position is obtained. Afterward, the energies as well as the gradients at arbitrary
expectation positions are obtained by cubic spline interpolation during the dynamics simulations.
We note that for multidimensional models and practical chemical systems,
this way of constructing CMES becomes computationally expensive. Fortunately, in practical
systems, CNEO-DFT
\cite{Xu20074106, Xu20084107}
minimizes the total energy of a system with nuclear expectation position
constraints and therefore can be used to calculate the CMES on the fly.
This work will focus more on building
the theoretical foundation and exploring the strengths and limitations of
CMES-MD with the help of the simple models.

In our practical model calculations,
classical MD and CMES-MD are performed with an in-house python script.
Note that throughout the Letter, by conventional MD or classical MD we mean
classical molecular simulations based on a Boltzmann sampling of the initial
velocity according to the designated temperature and evolving with classical
Newtonian equations. We are aware that other methods such as
quasi-classical MD
\cite{Karplus642033}
and SC-IVR
\cite{Wang989726, Sun987064, Liu09074113}
that can include zero-point effects in
classical simulations exist,
but we do not refer to them as classical MD or
conventional MD in this Letter.
The total simulation time of all conventional MD and CMES-MD simulations is chosen to
be 50\,ps, and the trajectories are integrated using the velocity-Verlet
algorithm with a time step of 0.5\,fs.
RPMD and CMD simulations on these one-dimensional models are performed with
a modified i-PI package.
\cite{Ceriotti141019}
Specifically, a 30\,ps PIMD trajectory with 0.1\,fs time step is first calculated
and used to generate initial configurations for RPMD and CMD. Then RPMD and CMD are
performed with a simulation length of 10\,ps. For RPMD, the time step is set to
0.1\,fs, and for CMD, the time step is 0.003125\,fs; the data are recorded
every 0.1\,fs.
For the Morse oscillator model, RPMD and CMD both use 64 beads in simulations
with $T=50\,\text{K}$, 32 beads in simulations with $T=300\,\text{K}$, and 16
beads in simulations with $T=1500\,\text{K}$. For the double-well potential
model, the corresponding bead numbers are 128 for $T=50\,\text{K}$, 64 for
$T=100\,\text{K}$, and 32 for $T=1500\,\text{K}$. We perform 1000 $NVE$
simulations whose initial configurations satisfy the Maxwell-Boltzmann
distribution to obtain an $NVT$ ensemble average for classical MD and CMES-MD;
however, this number is reduced to 50 for RPMD and 30 for CMD to limit
the computational cost.  After the simulations, the
trajectories are used to generate correlation functions. In data processing, a
correlation depth of 4096 points is used for MD and CMES-MD. For RPMD and CMD,
a larger depth of 16384 points is used because of the shorter time step used.
Power spectra are obtained via Fourier transforms of
the corresponding velocity autocorrelation functions. They are then averaged to
obtain the $NVT$ ensemble-averaged power spectra for each method. The intensities
of the averaged power spectra are finally adjusted so that they integrate to a
number that is proportional to the simulation temperature.

\begin{figure}[htpb]
    \centering
    \includegraphics[width=\linewidth]{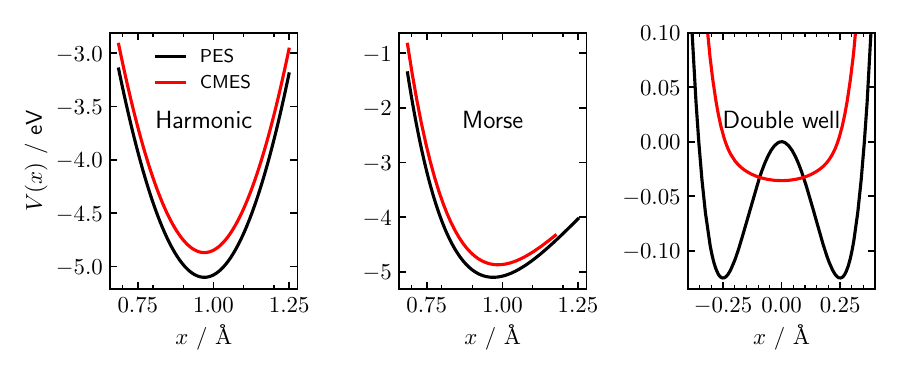}
    \caption{Comparison between the potential energy surface (PES)
    and constrained minimized energy surface (CMES) for the harmonic oscillator
    model (left), the Morse oscillator model (middle), and the double-well potential
    model (right).}
  \label{fig:potentials}
\end{figure}

The harmonic oscillator
$\hat{H} = \hat{p}^2 / 2 m + m \omega^2 \left( \hat{x} - x_{\text{e}} \right)^2 / 2$
is one particular model for which classical MD gives the same
trajectory as the exact quantum theory. CMD and RPMD are also exact for this
model system. For CMES-MD, because
$\hat{H}+f\hat{x}$
represents the harmonic oscillator with a shifted energy and a shifted
position based on the value of
$f$,
the constrained minimized energy state
$|A\rangle$
for any expectation position
$\langle \hat{x} \rangle=X$
is the ground state wave function of
$\hat{H}$
shifted to the expectation position
$\langle \hat{x} \rangle=X$.
Therefore, the corresponding energy surface as a function of expectation
position
$X$
is
\begin{equation}
  V^{\text{CMES}}(X)
  = \frac{\hbar \omega}{2} + \frac{1}{2} m \omega^2
  \left( X - x_{\text{e}} \right)^2.
\end{equation}
This effective potential universally shifts the original harmonic potential
upward by
$\hbar \omega / 2$,
which is the zero-point energy for a harmonic oscillator
(Figure~\ref{fig:potentials}). This result may seem counterintuitive because conventionally ZPE is
considered to be a property of the whole energy surface rather than a
point-wise property; however, we note that here the ZPE should be more
accurately considered as a quantum delocalization energy, which always exists
as the quantum wave packet travels through space. Because classical MD
produces the exact trajectory on the harmonic potential, the trajectory
produced by CMES-MD on
$V^{\text{CMES}}$
is also exact without any need for numerical tests.

\begin{figure}[htpb]
    \centering
    \includegraphics[width=\linewidth]{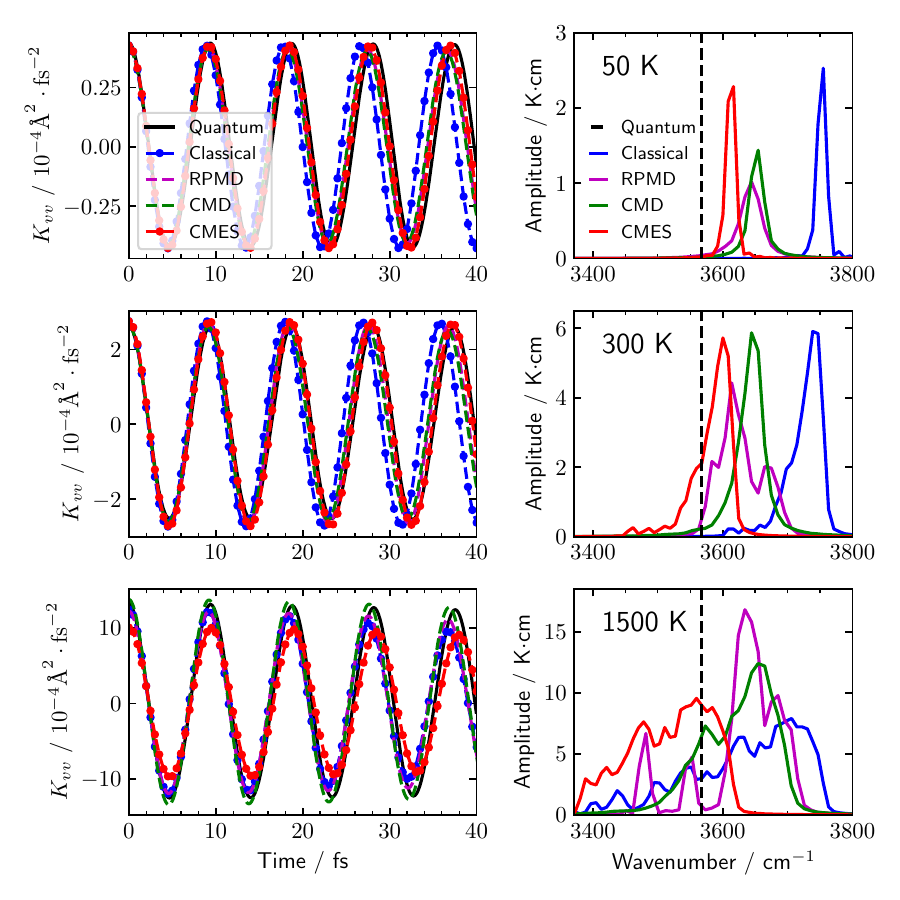}
    \caption{Velocity autocorrelation functions $K_{vv}$ and power spectra
    of the Morse oscillator mimicking the \ce{^16O^1H} radical
    at 50\,K (top), 300\,K (middle) and 1500\,K (bottom). The potential
    form is
    $V (x) = D_{\text{e}}
    ( \mathrm{e}^{ - 2 \alpha \left( x - x_{\text{e}} \right)}
    - 2 \mathrm{e}^{ - \alpha \left( x - x_{\text{e}} \right)} )$,
    where
    $D_{\text{e}} = h c \omega_{\text{e}}^2 / 4 \omega_{\text{e}} \chi_{\text{e}}$
    and
    $\alpha = \sqrt{2 \mu h c \omega_{\text{e}} \chi_{\text{e}} / \hbar^2}$.
    All of the parameters are the same as those used in
    ref~\citenum{Rossi14234116},
    where $\omega_{\text{e}}=3737.76\,\text{cm}^{-1}$,
    $\omega_{\text{e}}\chi_{\text{e}}=84.881\,\text{cm}^{-1}$,
    and $x_{\text{e}}=0.96966\,\text{\AA}$.
    The dashed vertical line represents the exact quantum frequency $3568\,\text{cm}^{-1}$.
    Two spikes below the quantum frequency in the RPMD 1500\,K spectrum are due to insufficient sampling.
    }
  \label{fig:morse}
\end{figure}

Compared with the harmonic potential, the Morse potential is a better model
for chemical bonds with anharmonic effects. Here we use a Morse potential that
can mimic the stretch of the
\ce{O-H}
radical and perform simulations using classical MD, CMES-MD, RPMD, and CMD.
Figure~\ref{fig:morse} shows the velocity autocorrelation functions and the
corresponding power spectra of these methods at three different temperatures.
The exact quantum results are used as references, which are
obtained from the analytical solution of the Morse potential.
Compared with the Kubo-transformed quantum velocity autocorrelation function,
\cite{Kubo57570}
classical MD underestimates the period of the correlation function and
therefore severely overestimates the vibrational frequency.
RPMD and CMD can more accurately describe the correlation function, and their
overestimations of the vibrational frequencies are significantly smaller.
CMES-MD has the best performance with good
agreement with the exact quantum correlation functions and more accurate vibrational frequencies.
The good results of CMES-MD at relatively low temperatures are not surprising because
CNEO-MD has been known to give accurate vibrational spectra at room temperature,
\cite{Xu224039}
and vibrational frequencies obtained from CNEO-DFT Hessian calculations are also
in great agreement with experimental values, indicating a good zero-temperature limit.
\cite{Xu21244110}
As the temperature increases, CMES-MD and other simulation methods start
to have broader and red-shifted peaks.
For this Morse potential, we observe that CMES-MD produces accurate spectra
for a temperature range between 50 and 1500\,K,
suggesting the reliability of CMES-MD
in the temperature range in which most chemical and biophysical reactions are performed.

\begin{figure}[htpb]
    \centering
    \includegraphics[width=\linewidth]{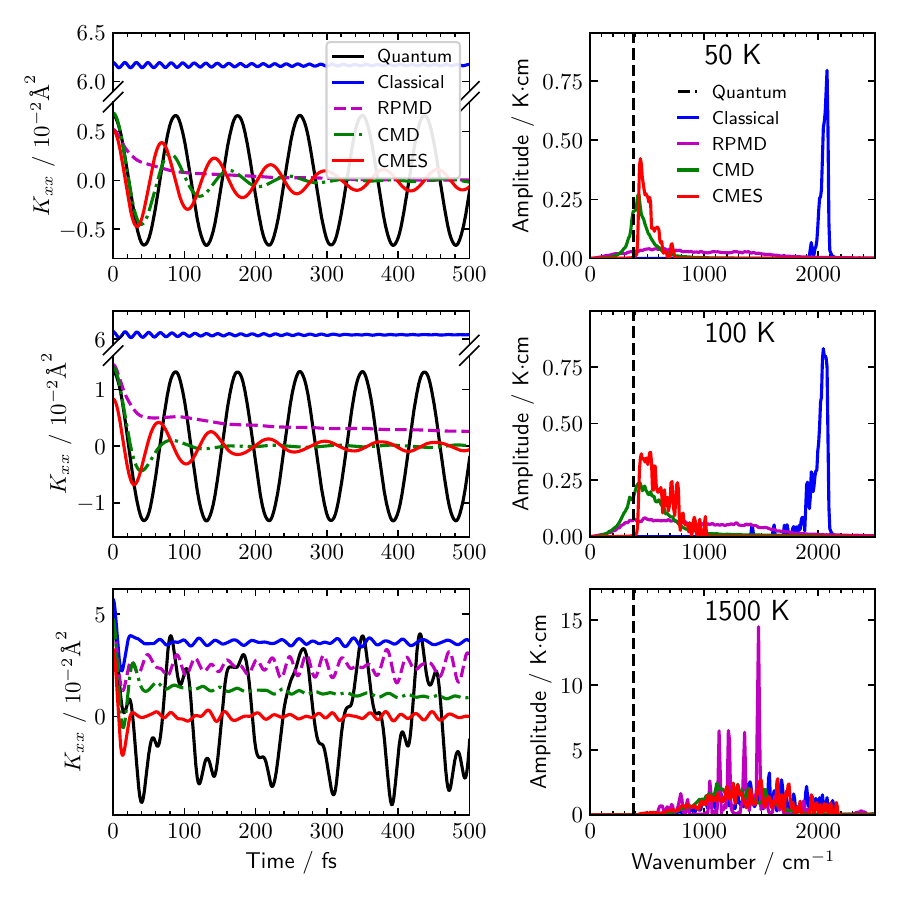}
    \caption{Position autocorrelation function $K_{xx}$ and power spectrum
    of a proton in the double-well potential at 50\,K (top),
    100\,K (middle) and 1500\,K (bottom). The potential form is
    $V (x) = a x^2 + b x^4$,
    where
    $a = -4 \,\text{eV} / \text{\AA}^2$
    and
    $b = 32 \,\text{eV} / \text{\AA}^4$.
    The dashed vertical line represents the exact quantum frequency $382\,\text{cm}^{-1}$.
    Multiple spikes in the RPMD 1500\,K spectrum are due to insufficient sampling.
    }
  \label{fig:dw}
\end{figure}

Next we investigate a more challenging double-well potential model, in which
quantum tunneling is expected to occur. We use a quartic double-well potential
with a
0.125$\,\text{eV}$
barrier height and a
0.5$\,\text{\AA}$
separation between the potential minima, which can roughly represent the
potential energy landscape for a practical proton transfer reaction.
As shown in Figure~\ref{fig:potentials}, the CMES of this double-well potential is a single well with the minimum
located at $X=0$ due to the symmetrical shape of the ground state wave
function with two peaks, whose expectation position is at $\langle \hat{x}\rangle =0$.
As the constrained expectation position deviates from the center,
the constrained minimized wave function becomes less symmetrical with more and more
excited state character mixed in, thus increasing the energy and forming a single-well
effective potential.
Note that this picture has also been observed
in the literature when investigating the zero-temperature limit of CMD.
\cite{Ramirez983303,Ramirez993339}

On this single-well effective potential, the quantum expectation position moves
smoothly between left and right as if the barrier does not exist. This is qualitatively
in agreement with the quantum picture, in which the wave function can tunnel back and forth
through the barrier with a smooth oscillation for the quantum expectation position.
This physical picture can be further quantitatively verified by the agreement
between the tunneling frequency by CMES-MD and the exact quantum tunneling frequency
(Figure~\ref{fig:dw}).
At low temperatures, classical MD simulations are all trapped in the local minima
of the double well and give position autocorrelation functions that are not vertically
centered at zero and a highly overestimated vibrational frequency that is close to
the second-order derivative at
the local minimum, indicating the failure of classical MD in describing tunneling effects.
The two path-integral methods show significant differences
in the double-well potential model, as CMD gives good autocorrelation functions
and predicts a relatively sharp peak with an accurate tunneling frequency,
whereas RPMD suffers from a fast decay of the correlation function \cite{Smith15244112} and a broad peak that
smears over a range of nearly 2000\,$\text{cm}^{-1}$.
CMES-MD is similar to CMD with a slightly overestimated
tunneling frequency. As the temperature increases,
classical MD shows red-shifts in the peak positions, and CMES-MD and CMD see blue-shifts.
At 1500\,K, all simulation methods behave very similarly with broad peaks that maximize around
1300-1500\,$\text{cm}^{-1}$.
All of these results show that CMES-MD performs reasonably well
in describing the dynamics in the double-well potential.

We note that quasi-classical MD
\cite{Karplus642033}
can be another way to include zero-point effects in
classical simulations and can describe well the vibrational
frequency of a Morse potential with an appropriate initial energy.
However, there are several known issues related to it.
For example, it can lead to zero-point energy leakage,
where the excess energy of one high-frequency mode may flow to a low-frequency one,
sometimes leading to unphysical molecular dissociations.
\cite{Guo96576, Ben-Nun968136}
Moreover, this method requires the calculation of harmonic frequencies to
approximate zero-point energies, which can be problematic if the potential is highly
anharmonic. One example is that in the double-well potential, the zero-point
energy obtained from the harmonic approximation will be highly inaccurate,
and like the classical MD case, the particle may still become trapped on either
side of the well if the classical barrier is higher than the zero-point energy.
In contrast, CMES-MD and path-integral based methods will not suffer from these problems.

In principle, the effective potential energy surfaces in MD simulations
should be temperature-dependent to fully account for nuclear quantum effects.
However, although CMES is a temperature-independent effective
potential, we observe good performance of CMES-MD over a relatively large
temperature range. This suggests that the adiabatic approximation that in essence
assumes that the quantum state adapts its wave function to the
lowest-energy state for a particular phase space point is reasonable. Nevertheless,
it is possible that when the
temperature is high and the particle is moving fast, the wave function
may not adapt fast enough to the constrained minimized wave function, thus breaking
the adiabatic approximation. Therefore, we expect CMES-MD to be more accurate at low
temperatures relative to the mode frequency.
Fortunately, for most vibrational modes, room temperature is still considered
low temperature, and therefore CMES-MD can be accurate in a good range of temperatures
typically investigated by chemical physicists and biophysicists.

Similar to conventional MD, the classical treatment brings not only
efficiency but also some limitations. For example, quantum coherence is
missing, which is reflected by a decreasing amplitude of the correlation
function (Figure~\ref{fig:dw}). Heat capacities will not approach zero when
$T\to 0\,\text{K}$
due to the loss of the energy quantization picture. Furthermore, classical
dynamics with distinguishable particles is incapable of capturing the exchange
effect, which is important in systems with heavily packed particles, such as in
a Bose-Einstein condensate.
\cite{Einstein05245, Bose24178}
Although detailed studies of these possible limitations are beyond the scope of this work,
they are important topics for our future research for better understanding the
applicability of CMES-MD.
We finally note that due to the similarity between CMES-MD and conventional classical MD,
in practical systems, we can expect analytical force field models or even
machine-learning force fields (ML-FFs) \cite{Unke2110142} to be built on the basis of the CMES,
which will allow for an even more efficient incorporation of NQEs in MD simulations.

In summary, we provide a new framework for incorporating NQEs
into classical molecular simulations. This is achieved
through the calculation of the CMES, which serves as the effective potential for
classical simulations. In CMES-MD, quantum delocalization and tunneling effects are
inherently included and therefore dynamical vibrational frequencies can be accurately described.
In simulations of practical systems,
CNEO-DFT can be used to obtain the CMES and the resulting CNEO-MD is
computationally more efficient than the popular RPMD and CMD methods when
{\it ab initio} PESs are used.
It may be further accelerated when combined with modern
machine-learning techniques in future developments.
As such, CMES/CNEO-MD is a promising new approach to describing NQEs in larger and
more complex systems, which will open the door to broader applications.

\begin{acknowledgement}
The authors thank David Manolopoulos, Stuart Althorpe, Edwin Sibert, Xi Xu, and
James Langford for helpful comments.
The authors are grateful for the funding support from the National Science Foundation
under Grant 2238473 and from the University of Wisconsin via
the Wisconsin Alumni Research Foundation.
\end{acknowledgement}

\begin{suppinfo}
Additional details, including
detailed derivations for the equations of motion
of CMES-MD for single-particle and multiple-particle cases, the connection
between CNEO-MD and CMES-MD, and a time propagation test on the performance
of CMES-MD.
\end{suppinfo}
%\nocite{*}

\bibliography{manuscript}
\end{document}

Here we present the detailed derivation for the equations of motion
of CMES-MD for single-particle and multiple-particle cases, as well as
its connection with CNEO-MD.
In the last section, as suggested by a reviewer, we show a test on
the performance of CMES-MD based on trajectory analysis.

\section{Detailed derivation for equations of motion of CMES-MD}

In this section, for simplicity, we assume that there is only one
quantum particle. However, the extension to the multi-particle case
is straightforward if the particles can be assumed to be distinguishable,
which will be shown in section \ref{sec:multi}.

\subsection{Decomposition of kinetic energy}

We start with the polar representation of a time-dependent wave function
\begin{equation}
\psi(\bm{x},t)=A(\bm{x},t)\exp\left(\frac{\mathrm{i}}{\hbar}S(\bm{x},t)\right),
\end{equation}
where $A$ and $S$ are both real functions. Then we can evaluate
a few quantities in terms of $A$ and $S$.

The expectation value of the position operator $\langle\hat{\bm{x}}\rangle(t)$
is
\begin{equation}
\langle\hat{\bm{x}}\rangle(t)=\langle\psi|\hat{\bm{x}}|\psi\rangle=\int\dd\bm{x}|\psi(\bm{x},t)|^{2}\bm{x}=\int\dd\bm{x}A^{2}(\bm{x},t)\bm{x},
\end{equation}

The expectation value of the momentum operator is
\begin{align}
\langle\hat{\bm{p}}\rangle(t) & =\langle\psi|\hat{\bm{p}}|\psi\rangle=-\mathrm{i}\hbar\int\dd\bm{x}\psi^{*}(\bm{x},t)\nabla\psi(\bm{x},t)\\
 & =-\mathrm{i}\hbar\int\dd\bm{x}A(\bm{x},t)\exp\left(-\frac{\mathrm{i}}{\hbar}S(\bm{x},t)\right)\nabla\left[A(\bm{x},t)\exp\left(\frac{\mathrm{i}}{\hbar}S(\bm{x},t)\right)\right]\\
 & =-\mathrm{i}\hbar\int\dd\bm{x}A(\bm{x},t)\nabla A(\bm{x},t)+\int\dd\bm{x}A^{2}(\bm{x},t)\nabla S(\bm{x},t)\\
 & =-\frac{\mathrm{i}\hbar}{2}\int\dd\bm{x}\nabla A^{2}(\bm{x},t)+\int\dd\bm{x}A^{2}(\bm{x},t)\nabla S(\bm{x},t)\\
 & =\int\dd\bm{x}A^{2}(\bm{x},t)\nabla S(\bm{x},t).\label{eqn:momentum}
\end{align}
Note that $\int\dd\bm{x}\nabla A^{2}(\bm{x},t)$ is zero because the
wave function needs to vanish at infinity.

Using the time-dependent Schr\"odinger equation
\begin{equation}
\mathrm{i}\hbar\frac{\partial}{\partial t}\psi(\bm{x},t)=\hat{H}(t)\psi(\bm{x},t),
\end{equation}
where the Hamiltonian operator consists of kinetic energy part and
potential energy part
\begin{equation}
\hat{H}(t)=-\frac{\hbar^{2}}{2m}\nabla^{2}+V(\bm{x},t).
\end{equation}
The first equation of the Ehrenfest theorem can be derived:
\begin{align}
\frac{\dd\langle\hat{\bm{x}}\rangle(t)}{\dd t} & =\int\dd\bm{x}\left[\frac{\partial\psi^{*}(\bm{x},t)}{\partial t}\psi(\bm{x},t)+\psi^{*}(\bm{x},t)\frac{\partial\psi(\bm{x},t)}{\partial t}\right]\bm{x}\\
 & =-\frac{\mathrm{i}}{\hbar}\int\dd\bm{x}[-\psi(\bm{x},t)\hat{H}\psi^{*}(\bm{x},t)+\psi^{*}(\bm{x},t)\hat{H}\psi(\bm{x},t)]\bm{x}\\
 & =\frac{\mathrm{i}\hbar}{2m}\int\dd\bm{x}[-\psi(\bm{x},t)\nabla^{2}\psi^{*}(\bm{x},t)+\psi^{*}(\bm{x},t)\nabla^{2}\psi(\bm{x},t)]\bm{x}\\
 & =\frac{\mathrm{i}\hbar}{2m}\int\dd\bm{x}[-\bm{x}\psi(\bm{x},t)\nabla^{2}\psi^{*}(\bm{x},t)+\bm{x}\psi^{*}(\bm{x},t)\nabla^{2}\psi(\bm{x},t)]\\
 & =\frac{\mathrm{i}\hbar}{2m}\int\dd\bm{x}\nabla[\bm{x}\psi(\bm{x},t)]\nabla\psi^{*}(\bm{x},t)-\nabla[\bm{x}\psi^{*}(\bm{x},t)]\nabla\psi(\bm{x},t)\\
 & =\frac{\mathrm{i}\hbar}{2m}\int\dd\bm{x}\big[\bm{x}\nabla\psi(\bm{x},t)\nabla\psi^{*}(\bm{x},t)+\psi(\bm{x},t)\nabla\psi^{*}(\bm{x},t)\nonumber \\
 & \quad-\bm{x}\nabla\psi^{*}(\bm{x},t)\nabla\psi(\bm{x},t)-\psi^{*}(\bm{x},t)\nabla\psi(\bm{x},t)\big]\\
 & =\frac{\mathrm{i}\hbar}{2m}\int\dd\bm{x}[\psi(\bm{x},t)\nabla\psi^{*}(\bm{x},t)-\psi^{*}(\bm{x},t)\nabla\psi(\bm{x},t)]\\
 & =\frac{1}{2m}\int\dd\bm{x}[\psi(\bm{x},t)(\mathrm{i}\hbar\nabla)\psi^{*}(\bm{x},t)+\psi^{*}(\bm{x},t)(-\mathrm{i}\hbar\nabla)\psi(\bm{x},t)]\\
 & =\frac{1}{m}\int\dd\bm{x}A^{2}(\bm{x},t)\nabla S(\bm{x},t)\\
 & =\frac{\langle\hat{\bm{p}}\rangle(t)}{m},\label{eqn:ehrenfest1}
\end{align}
where the potential parts of the Hamiltonian cancel, and the we have
used again the requirement that the wave function needs to vanish
at infinity.

Now we derive eq~1 in the main paper, which is the kinetic energy
in terms of $A$ and $S$:
\begin{align}
\langle\hat{T}\rangle(t) & =\frac{\langle\hat{\bm{p}}^{2}\rangle(t)}{2m}\\
 & =-\frac{\hbar^{2}}{2m}\int\dd\bm{x}\psi^{*}(\bm{x},t)\nabla^{2}\psi(\bm{x},t)\\
 & =\frac{\hbar^{2}}{2m}\int\dd\bm{x}[\nabla\psi^{*}(\bm{x},t)]\cdot[\nabla\psi(\bm{x},t)]\\
 & =\frac{\hbar^{2}}{2m}\int\dd\bm{x}\left\{ \nabla\left[A(\bm{x},t)\exp\left(-\frac{\mathrm{i}}{\hbar}S(\bm{x},t)\right)\right]\right\} \cdot\left\{ \nabla\left[A(\bm{x},t)\exp\left(\frac{\mathrm{i}}{\hbar}S(\bm{x},t)\right)\right]\right\} \\
 & =\frac{\hbar^{2}}{2m}\int\dd\bm{x}\left\{ [\nabla A(\bm{x},t)]\exp\left(-\frac{\mathrm{i}}{\hbar}S(\bm{x},t)\right)-\frac{\mathrm{i}}{\hbar}[\nabla S(\bm{x},t)]A(\bm{x},t)\exp\left(-\frac{\mathrm{i}}{\hbar}S(\bm{x},t)\right)\right\} \cdot\nonumber \\
 & \quad\left\{ [\nabla A(\bm{x},t)]\exp\left(\frac{\mathrm{i}}{\hbar}S(\bm{x},t)\right)+\frac{\mathrm{i}}{\hbar}[\nabla S(\bm{x},t)]A(\bm{x},t)\exp\left(\frac{\mathrm{i}}{\hbar}S(\bm{x},t)\right)\right\} \\
 & =\frac{\hbar^{2}}{2m}\int\dd\bm{x}[\nabla A(\bm{x},t)]^{2}+\frac{1}{2m}\int\dd\bm{x}A^{2}(\bm{x},t)[\nabla S(\bm{x},t)]^{2}\\
 & =\int\dd\bm{x}A(\bm{x},t)\frac{(-\mathrm{i}\hbar\nabla)^{2}}{2m}A(\bm{x},t)+\frac{1}{2m}\int\dd\bm{x}A^{2}(\bm{x},t)[\nabla S(\bm{x},t)]^{2}.
\label{eqn:kineticEnergyNoSigma}
\end{align}
Here the integration by parts is used again with wave function boundary
conditions. We can see that the kinetic energy can be decomposed into
two parts.

The first part is only about amplitude $A$, which roughly depends
on how localized $A$ is, and is the zero-point kinetic energy if
phase part $S$ does not exist.

The second part is directly about $\nabla S$, which is related to
the observable momentum. In fact, with the definition of the momentum
field $p(\bm{x},t)\equiv\nabla S(\bm{x},t)$ according to Bohmian
mechanics,\cite{de1926possibilite,de1927structure,Madelung27322,Bohm52166,Bohm52180,schleich2013schrodinger}
this term can be written as
\begin{equation}
\frac{1}{2m}\int\dd\bm{x}A^{2}(\bm{x},t)[\nabla S(\bm{x},t)]^{2}=\frac{\langle\hat{\bm{p}}\rangle^{2}(t)}{2m}+\frac{\sigma_{\bm{p}}^{2}(t)}{2m},
\end{equation}
where $\sigma_{\bm{p}}^{2}(t)$ is the variance of the observable
momentum field that is defined as
\begin{equation}
\sigma_{\bm{p}}^{2}(t)\equiv\int\dd\bm{x}A^{2}(\bm{x},t)[\nabla S(\bm{x},t)]^{2}-\langle\hat{\bm{p}}\rangle^{2}(t).
\end{equation}
In this way, the total kinetic energy can be written as
\begin{equation}
\langle\hat{T}\rangle(t)=\langle A(t)|\hat{T}|A(t)\rangle+\frac{\langle\hat{\bm{p}}\rangle^{2}(t)}{2m}+\frac{\sigma_{\bm{p}}^{2}(t)}{2m}.
\label{eqn:kineticEnergy}
\end{equation}
We can see that with this decomposition, the total kinetic energy
is the sum of: (1) the zero-point kinetic energy from the delocalized
amplitude part, (2) the classical kinetic energy from the observable
momentum, and (3) the variance of the observable momentum field.

\subsection{Equations of motion}

Equation~\ref{eqn:ehrenfest1} gives the $\dd\bm{x}/\dd t$ part of equation
of motion. In order to get the $\dd\bm{p}/\dd t$ part of the equations
of motion, we start with the following energy identity,
\begin{equation}
\langle\hat{H}\rangle=\langle\hat{T}\rangle+\langle\hat{V}\rangle.
\end{equation}
Then taking the time derivative on each term gives
\begin{align}
\frac{\dd}{\dd t}\langle\hat{T}\rangle & =\frac{\dd}{\dd t}\langle\hat{H}\rangle-\frac{\dd}{\dd t}\langle\hat{V}\rangle\\
 & =\langle\frac{\partial\hat{H}}{\partial t}\rangle-\frac{\dd}{\dd t}\langle\hat{V}\rangle\\
 & =\langle\frac{\partial V(\bm{x},t)}{\partial t}\rangle-\frac{\dd}{\dd t}\langle\hat{V}\rangle.\label{eqn:TeqHmV}
\end{align}
Here the first step has used the time-dependent Schr\"odinger equation,
and the second step has used that the kinetic energy operator is not
time-dependent. We note that in many systems of interest, the external
potential part is also time-independent, however, we keep its time
dependence to account for those cases where the external potential
changes with time, for example, a time-dependent electric field. Later
we will show that the final conclusion does not change whether $V$
is time-dependent or not because $\langle\frac{\partial V(\bm{x},t)}{\partial t}\rangle$
will get canceled.

Taking the time derivative of eq~\ref{eqn:kineticEnergy} and combining
with eq~\ref{eqn:TeqHmV}, we have
\begin{equation}
\frac{\dd}{\dd t}\langle A(t)|\hat{T}|A(t)\rangle+\frac{\dd}{\dd t}\frac{\langle\hat{\bm{p}}\rangle^{2}}{2m}+\frac{\dd}{\dd t}\frac{\sigma_{\bm{p}}^{2}}{2m}=\langle\frac{\partial V(\bm{x},t)}{\partial t}\rangle-\frac{\dd}{\dd t}\langle\hat{V}\rangle.\label{eqn:dTdt}
\end{equation}
Using $\frac{\dd}{\dd t}\frac{\langle\hat{\bm{p}}\rangle^{2}}{2m}=\frac{\langle\hat{\bm{p}}\rangle}{m}\cdot\frac{\dd\langle\hat{\bm{p}}\rangle}{\dd t}$,
we can move most terms to the right hand side of eq~\ref{eqn:dTdt}
and get
\begin{align}
\frac{\langle\hat{\bm{p}}\rangle}{m}\cdot\frac{\dd\langle\hat{\bm{p}}\rangle}{\dd t} & =\langle\frac{\partial V}{\partial t}\rangle-\frac{\dd}{\dd t}\langle A(t)|\hat{T}|A(t)\rangle-\frac{\dd}{\dd t}\langle A(t)|\hat{V}|A(t)\rangle-\frac{\dd}{\dd t}\frac{\sigma_{\bm{p}}^{2}}{2m}\\
 & =\langle\frac{\partial V}{\partial t}\rangle-\frac{\dd}{\dd t}\langle A(t)|\hat{H}|A(t)\rangle-\frac{\dd}{\dd t}\frac{\sigma_{\bm{p}}^{2}}{2m}.\label{eqn:exactEOM}
\end{align}
The derivation up to here is exact with no approximation imposed.
However, this equation for the $\dd\bm{p}/\dd t$ part is not practically
useful as the time derivatives on the right-hand side are unknown.

\subsection{Constrained minimization approximation}

In order to make eq~\ref{eqn:exactEOM} an equation of motion that
can be practically used in molecular dynamics simulations, here we
proceed with an approximation that builds a connection between quantum
states and the classical phase space.

Conventionally, if the potential is slowly varying in space, the Ehrenfest
theorem provides a connection between the classical Newtonian dynamics
in phase space $(\bm{X},\bm{P})$ and the evolution of quantum
expectation position and momentum $(\langle\hat{\bm{x}}\rangle,\langle\hat{\bm{p}}\rangle)$.
Here with the same goal of mapping quantum states with the classical
phase space, we approximate the quantum state as the energy-minimized
state for a given classical phase space point. That is, when the system
is at a particular phase space point given by an expectation position
and an expectation momentum, i.e., $(\langle\hat{\bm{x}}\rangle,\langle\hat{\bm{p}}\rangle)=(\bm{X},\bm{P})$,
we assume that quantum state $|\Psi\rangle$ always adapts to
the energy-minimized state for that phase space point. This approximation
is in essence an adiabatic approximation.

Under this approximation, quantum state $|\Psi\rangle$ becomes
an explicit function of $(\bm{X},\bm{P})$ and an implicit function
of time $t$, i.e., $|\Psi\rangle(\bm{X}(t),\bm{P}(t))$. At a particular
phase space point $(\bm{X}(t),\bm{P}(t))$ that the system evolves
to at time $t$, the state can be obtained with a constrained energy
minimization procedure. We write the Lagrangian as
\begin{equation}
\mathcal{L}=\langle\psi|\hat{H}(t)|\psi\rangle+\bm{f}\cdot(\langle\psi|\hat{\bm{x}}|\psi\rangle-\bm{X}(t))-\bm{v}\cdot(\langle\psi|\hat{\bm{p}}|\psi\rangle-\bm{P}(t))-\tilde{E}(\langle\psi|\psi\rangle-1).
\end{equation}
where $\bm{f}$ is the Lagrange multiplier associated with the expectation
position, $\bm{v}$ is the Lagrange multiplier associated with the
expectation momentum, and $\tilde{E}$ is the Lagrange multiplier
associated with the wave function normalization. We can further express
the Lagrangian in terms of $A$ and $S$, according to eqs~\ref{eqn:momentum}
and \ref{eqn:kineticEnergyNoSigma}:
\begin{align}
\mathcal{L} & =\langle\psi|\hat{T}+\hat{V}|\psi\rangle+\bm{f}\cdot(\langle A|\hat{\bm{x}}|A\rangle-\bm{X}(t))\nonumber \\
 & \quad-\bm{v}\cdot\left(\int\dd\bm{x}A^{2}(\bm{x})\nabla S(\bm{x})-\bm{P}(t)\right)-\tilde{E}(\langle\psi|\psi\rangle-1)\\
 & =\langle A|\hat{V}|A\rangle+\langle A|\hat{T}|A\rangle+\frac{1}{2m}\int\dd\bm{x}A^{2}(\bm{x})[\nabla S(\bm{x})]^{2}+\bm{f}\cdot(\langle A|\hat{\bm{x}}|A\rangle-\bm{X}(t))\nonumber \\
 & \quad-\bm{v}\cdot\left(\int\dd\bm{x}A^{2}(\bm{x})\nabla S(\bm{x})-\bm{P}(t)\right)-\tilde{E}(\langle A|A\rangle-1)\\
 & =\langle A|\hat{H}|A\rangle+\frac{1}{2m}\int\dd\bm{x}A^{2}(\bm{x})[\nabla S(\bm{x})]^{2}+\bm{f}\cdot(\langle A|\hat{\bm{x}}|A\rangle-\bm{X}(t))\nonumber \\
 & \quad-\bm{v}\cdot\left(\int\dd\bm{x}A^{2}(\bm{x})\nabla S(\bm{x})-\bm{P}(t)\right)-\tilde{E}(\langle A|A\rangle-1).
\end{align}
At the state that minimizes the energy, the Lagrangian should be stationary
with respect to $A$, $S$, as well as all the Lagrange multipliers.

We first take the functional derivative with respect to $S$ part,
\begin{equation}
\frac{\delta\mathcal{L}}{\delta[\nabla S]}=A^{2}(\bm{x})\left[\frac{\nabla S(\bm{x})}{m}-\bm{v}\right],
\end{equation}
and when it equals to zero, we have
\begin{equation}
\nabla S(\bm{x})=m\bm{v}.
\end{equation}
This result means that under the adiabatic approximation, the momentum
field is trivially uniform. Thus, according to the definition of the
observable momentum field variance
\begin{equation}
\sigma_{\bm{p}}^{2}=\int\dd\bm{x}A^{2}(\bm{x},t)[\nabla S(\bm{x},t)]^{2}-
\left[\int\dd\bm{x}A^{2}(\bm{x},t)\nabla S(\bm{x},t)\right]^{2}=(m\bm{v})^{2}-(m\bm{v})^{2}=0.
\end{equation}

Then the requirement on momentum constraint gives
\begin{align}
\bm{P}(t) & =\langle\hat{\bm{p}}\rangle(t)\\
 & =\int\dd\bm{x}A^{2}(\bm{x},t)\nabla S(\bm{x},t)\\
 & =m\bm{v},
\end{align}
which further leads to the solution of the lagrange multiplier $\bm{v}$
\begin{equation}
\bm{v}=\frac{\bm{P}(t)}{m}.
\end{equation}
It is nothing but the velocity. Thus the momentum field as well as
the phase part of the wave function is analytically solved and it
is uniformly being
\begin{equation}
\nabla S(\bm{x})=\bm{P}(t).
\end{equation}

Next we take the functional derivative for amplitude $A$ part:
\begin{equation}
\frac{\delta\mathcal{L}}{\delta A}=2\hat{H}A+\frac{1}{m}A(\nabla S)^{2}+2\bm{f}\cdot\hat{\bm{x}}A-2\bm{v}\cdot\nabla SA-2\tilde{E}A.
\end{equation}
Making it zero leads to the eigenvalue equation
\begin{equation}
\left[\hat{H}(t)+\frac{(\nabla S)^{2}}{2m}+\bm{f}\cdot\hat{\bm{x}}-\bm{v}\cdot\nabla S\right]|A\rangle=\tilde{E}|A\rangle.
\end{equation}
Because by solving the phase part we have known that $\nabla S=\bm{P}$
and $\bm{v}=\bm{P}/m$, this eigenvalue equation can further reduce
to
\begin{equation}
[\hat{H}(t)+\bm{f}\cdot\hat{\bm{x}}]|A\rangle=\left(\tilde{E}+\frac{\bm{P}^{2}}{2m}\right)|A\rangle.
\end{equation}
The eigenvalue $\tilde{E}+\bm{P}^{2}/2m$, eigenstate $|A\rangle$
, and the Lagrange multiplier $\bm{f}$ can be solved under the expectation
position and normalization constraints for $|A\rangle$. Note again
that the momentum constraint part has been solved analytically above.

With the adiabatic approximation, the solution of $A$ only depends
on current position $\bm{X}$, therefore, $A$ is an explicit
function of $\bm{X}$ and an implicit function of time $t$. Thus,
the variable dependence can be written as:
\begin{equation}
A(\bm{x},t)=A(\bm{x};\bm{X}(t)).
\end{equation}
Thus, with the adiabatic approximation, we apply the chain rule to
eq~\ref{eqn:exactEOM} and obtain
\begin{align}
\frac{\langle\hat{\bm{p}}\rangle}{m}\cdot\frac{\dd\langle\hat{\bm{p}}\rangle}{\dd t} & \approx\langle\frac{\partial V}{\partial t}\rangle-\frac{\dd}{\dd t}\langle A(\bm{X}(t))|\hat{H}|A(\bm{X}(t))\rangle\\
 & =\langle\frac{\partial V}{\partial t}\rangle-\langle A(\bm{X}(t))|\frac{\partial}{\partial t}\hat{H}|A(\bm{X}(t))\rangle-\langle\frac{\dd}{\dd t}A(\bm{X}(t))|\hat{H}|A(\bm{X}(t))\rangle\nonumber \\
 & \quad-\langle A(\bm{X}(t))|\hat{H}|\frac{\dd}{\dd t}A(\bm{X}(t))\rangle\\
 & =\langle\frac{\partial V}{\partial t}\rangle-\langle\frac{\partial V}{\partial t}\rangle-\langle\frac{\dd}{\dd t}A(\bm{X}(t))|\hat{H}|A(\bm{X}(t))\rangle-\langle A(\bm{X}(t))|\hat{H}|\frac{\dd}{\dd t}A(\bm{X}(t))\rangle\\
 & =-\langle\frac{\dd}{\dd t}A(\bm{X}(t))|\hat{H}|A(\bm{X}(t))\rangle-\langle A(\bm{X}(t))|\hat{H}|\frac{\dd}{\dd t}A(\bm{X}(t))\rangle\\
 & =-\frac{\dd\bm{X}(t)}{\dd t}\cdot[\langle\nabla_{\bm{X}}A|\hat{H}|A\rangle+\langle A|\hat{H}|\nabla_{\bm{X}}A\rangle]\\
 & =-\frac{\langle\hat{\bm{p}}\rangle}{m}\cdot\nabla_{\bm{X}}\langle A|\hat{H}|A\rangle.
\end{align}
Note that $\hat{H}$ is dependent on $\bm{x}$, but it has no dependence
on expectation position $\bm{X}$. Another interesting thing is
that $\langle\frac{\partial V}{\partial t}\rangle$ gets canceled
and does not explicitly show in the final result. Finally, according
to classical mechanics, it is natural to assume that the change of
$\langle\hat{\bm{p}}\rangle$ should have an opposite direction to
the energy gradient term ($\nabla_{\bm{X}}\langle A|\hat{H}|A\rangle$),
therefore, the common prefactor $|\langle\hat{\bm{p}}\rangle|/m$
can be dropped, and we arrive at the final expression
\begin{equation}
\frac{\dd\langle\hat{\bm{p}}\rangle}{\dd t}\approx-\nabla_{\bm{X}}\langle A|\hat{H}|A\rangle.\label{eqn:modehrenfest}
\end{equation}
This equation is simple and elegant, but we note that it should not
be surprising to see this result because this equation essentially
allows the energy transfer between observable kinetic energy and effective
potential energy (including both regular potential energy and zero-point
energy) with a conservation of total energy.

Equations \ref{eqn:ehrenfest1} and \ref{eqn:modehrenfest} together make
the equations of motion for CMES-MD.

\section{Generalization to multiple particles}

\label{sec:multi}

Here we generalize the CMES-MD idea to the multi-particle case. The
idea is completely the same as the section above but with slightly
more tedious derivations.

Assume now we have $N$ particles that are distinguishable (Note:
we have to assume them to be distinguishable because if they are indistinguishable,
the positions for each individual particles cannot be defined), and
the potential energy surface $V(\bm{x}_{1},\cdots,\bm{x}_{N})$ is
also known. The total wave function for these $N$ particles should
be a correlated form, but here we use a mean-field approximation and
write the total nuclear wave function as a Hartree product:
\begin{equation}
\Psi(\bm{x}_{1},\cdots,\bm{x}_{N},t)=\psi_{1}(\bm{x}_{1},t)\cdots\psi_{N}(\bm{x}_{N},t).
\end{equation}
Each wave function $\psi_{i}$ has its polar representation:
\begin{equation}
\psi_{i}(\bm{x}_{i},t)=A_{i}(\bm{x}_{i},t)\exp\left(\frac{\mathrm{i}}{\hbar}S(\bm{x}_{i},t)\right).
\end{equation}
Their positions and momenta are
\begin{align}
\langle\hat{\bm{x}}_{i}\rangle(t) & =\int\dd\bm{x}_{1}\cdots\dd\bm{x}_{N}|\Psi(\bm{x}_{1},\cdots,\bm{x}_{N},t)|^{2}\bm{x}_{i}\\
 & =\int\dd\bm{x}_{1}\cdots\dd\bm{x}_{j\neq i}\cdots\dd\bm{x}_{N}A_{1}^{2}(\bm{x}_{1},t)\cdots A_{j\neq i}^{2}(\bm{x}_{j\neq i},t)\cdots A_{N}^{2}(\bm{x}_{N},t)\int\dd\bm{x}_{i}A_{i}^{2}(\bm{x}_{i},t)\bm{x}_{i}\\
 & =\int\dd\bm{x}_{i}A_{i}^{2}(\bm{x}_{i},t)\bm{x}_{i},
\end{align}
\begin{align}
\langle\hat{\bm{p}}_{i}\rangle(t) & =-\mathrm{i}\hbar\int\dd\bm{x}_{1}\cdots\dd\bm{x}_{N}\Psi^{\ast}(\bm{x}_{1},\cdots,\bm{x}_{N},t)\nabla_{i}\Psi(\bm{x}_{1},\cdots,\bm{x}_{N},t)\\
 & =\int\dd\bm{x}_{1}\cdots\dd\bm{x}_{j\neq i}\cdots\dd\bm{x}_{N}A_{1}^{2}(\bm{x}_{1},t)\cdots A_{j\neq i}^{2}(\bm{x}_{j\neq i},t)\cdots A_{N}^{2}(\bm{x}_{N},t)\times\nonumber \\
 & \quad\int\dd\bm{x}_{i}A_{i}^{2}(\bm{x}_{i},t)\nabla_{i}S_{i}(\bm{x}_{i},t)\\
 & =\int\dd\bm{x}_{i}A_{i}^{2}(\bm{x}_{i},t)\nabla_{i}S_{i}(\bm{x}_{i},t).
\end{align}
The total kinetic energy is
\begin{align}
\langle\hat{T}\rangle(t) & =\sum_{i}^{N}\langle\hat{T}_{i}\rangle(t)\\
 & =\sum_{i}^{N}\frac{\langle\hat{\bm{p}}_{i}^{2}\rangle(t)}{2m_{i}}\\
 & =\sum_{i}^{N}-\frac{\hbar^{2}}{2m_{i}}\int\dd\bm{x}_{1}\cdots\dd\bm{x}_{N}\Psi^{\ast}(\bm{x}_{1},\cdots,\bm{x}_{N},t)\nabla_{i}^{2}\Psi(\bm{x}_{1},\cdots,\bm{x}_{N},t)\\
 & =\sum_{i}^{N}\frac{\hbar^{2}}{2m_{i}}\int\dd\bm{x}_{i}[\nabla_{i}\psi_{i}^{\ast}(\bm{x}_{i},t)]\cdot[\nabla_{i}\psi_{i}(\bm{x}_{i},t)]\\
 & =\sum_{i}^{N}\int\dd\bm{x}_{i}A_{i}(\bm{x}_{i},t)\frac{(-\mathrm{i}\hbar\nabla_{i})^{2}}{2m_{i}}A_{i}(\bm{x}_{i},t)+\frac{1}{2m_{i}}\int\dd\bm{x}_{i}A_{i}^{2}(\bm{x}_{i},t)[\nabla_{i}S_{i}(\bm{x}_{i},t)]^{2}\\
 & =\sum_{i}^{N}\langle A_{i}(t)|\hat{T}_{i}|A_{i}(t)\rangle+\frac{\langle\hat{\bm{p}}_{i}\rangle^{2}(t)}{2m_{i}}+\frac{\sigma_{\bm{p}_{i}}^{2}(t)}{2m_{i}}.\label{eqn:kineticMulti}
\end{align}
Similar to the single-particle case, we have the time derivative of
energy identity:
\begin{equation}
\frac{\dd}{\dd t}\langle\hat{T}\rangle=\frac{\dd}{\dd t}\langle\hat{H}\rangle-\frac{\dd}{\dd t}\langle\hat{V}\rangle=\langle\frac{\partial V(\bm{x}_{1},\cdots,\bm{x}_{N},t)}{\partial t}\rangle-\frac{\dd}{\dd t}\langle\hat{V}(\bm{x}_{1},\cdots,\bm{x}_{N},t)\rangle.\label{eqn:TeqHmVMulti}
\end{equation}
It further gives
\begin{align}
\sum_{i}^{N}\frac{\langle\hat{\bm{p}}_{i}\rangle}{m_{i}}\cdot\frac{\dd\langle\hat{\bm{p}}_{i}\rangle}{\dd t} & =\langle\frac{\partial V(\bm{x}_{1},\cdots,\bm{x}_{N},t)}{\partial t}\rangle-\frac{\dd}{\dd t}\langle\hat{V}(\bm{x}_{1},\cdots,\bm{x}_{N},t)\rangle-\nonumber \\
 & \quad\sum_{i}^{N}\left\{ \frac{\dd}{\dd t}\langle A_{i}(t)|\hat{T}_{i}|A_{i}(t)\rangle+\frac{\dd}{\dd t}\frac{\sigma_{\bm{p}_{i}}^{2}(t)}{2m_{i}}\right\} \\
 & =\langle\frac{\partial V(\bm{x}_{1},\cdots,\bm{x}_{N},t)}{\partial t}\rangle-\frac{\dd}{\dd t}\langle A_{1}\cdots A_{N}|\hat{T}+V(t)|A_{1}\cdots A_{N}\rangle-\sum_{i}^{N}\frac{\dd}{\dd t}\frac{\sigma_{\bm{p}_{i}}^{2}(t)}{2m_{i}}.\label{eqn:multiExact}
\end{align}
The Lagrangian for the multi-particle case is
\begin{align}
\mathcal{L} & =\langle\Psi|\hat{H}(t)|\Psi\rangle+\sum_{i}^{N}\bm{f}_{i}\cdot(\langle\hat{\bm{x}}_{i}\rangle-\bm{X}_{i}(t))-\bm{v}_{i}\cdot(\langle\hat{\bm{p}}_{i}\rangle-\bm{P}_{i}(t))-\tilde{E}_{i}(\langle\psi_{i}|\psi_{i}\rangle-1)\\
 & =\langle A_{1}\cdots A_{N}|\hat{V}(t)|A_{1}\cdots A_{N}\rangle+\sum_{i}^{N}\langle A_{i}|\hat{T}_{i}|A_{i}\rangle+\frac{1}{2m_{i}}\int\dd\bm{x}_{i}A_{i}^{2}(\bm{x}_{i})[\nabla_{i}S_{i}(\bm{x}_{i})]^{2}\nonumber \\
 & \quad+\bm{f}_{i}\cdot(\langle A_{i}|\hat{\bm{x}}_{i}|A_{i}\rangle-\bm{X}_{i}(t))-\bm{v}_{i}\cdot\left(\int\dd\bm{x}_{i}A_{i}^{2}(\bm{x}_{i})\nabla_{i}S_{i}(\bm{x}_{i})-\bm{P}_{i}(t)\right)-\tilde{E}_{i}(\langle A_{i}|A_{i}\rangle-1).
\end{align}
Taking the functional derivative with respect to $\nabla_{i}S_{i}(\bm{x}_{i})$
gives
\begin{equation}
\frac{\delta\mathcal{L}}{\delta[\nabla_{i}S_{i}]}=A_{i}^{2}(\bm{x}_{i})\left[\frac{\nabla_{i}S_{i}(\bm{x}_{i})}{m_{i}}-\bm{v}_{i}\right].
\end{equation}
Making it zero leads to the solution
\begin{equation}
\nabla_{i}S_{i}(\bm{x}_{i})=m_{i}\bm{v}_{i}=\bm{P}_{i}(t),
\end{equation}
which also makes each individual $\sigma_{\bm{p}_{i}}^{2}$ zero.

Taking the functional derivative with respect to $A_{i}(\bm{x}_{i})$
gives
\begin{align}
\frac{\delta\mathcal{L}}{\delta A_{i}} & =2A_{i}\int\dd\bm{x}_{1}\cdots\dd\bm{x}_{j\neq i}\cdots\dd\bm{x}_{N}A_{1}^{2}(\bm{x}_{1})\cdots A_{j\neq i}^{2}(\bm{x}_{j\neq i})\cdots A_{N}^{2}(\bm{x}_{N})V(\bm{x}_{1},\cdots,\bm{x}_{N},t)\nonumber \\
 & \quad+2\hat{T}_{i}A_{i}+\frac{1}{m_{i}}A_{i}(\nabla_{i}S_{i})^{2}+2\bm{f}_{i}\cdot\hat{\bm{x}}_{i}A_{i}-2\bm{v_{i}}\cdot\nabla_{i}S_{i}A_{i}-2\tilde{E}_{i}A_{i}\\
 & =2V_{i}(\bm{x}_{i},t)A_{i}+2\hat{T}_{i}A_{i}+2\bm{f}_{i}\cdot\hat{\bm{x}}_{i}A_{i}-2\left(\tilde{E}_{i}+\frac{\bm{P}_{i}^{2}}{2m_{i}}\right)A_{i},\label{eqn:multiA}
\end{align}
where $V_{i}$ is the effective potential that particle $i$ is experiencing:
\begin{equation}
V_{i}(\bm{x}_{i},t)\equiv\int\dd\bm{x}_{1}\cdots\dd\bm{x}_{j\neq i}\cdots\dd\bm{x}_{N}A_{1}^{2}(\bm{x}_{1})\cdots A_{j\neq i}^{2}(\bm{x}_{j\neq i})\cdots A_{N}^{2}(\bm{x}_{N})V(\bm{x}_{1},\cdots,\bm{x}_{N},t).
\end{equation}
Making eq~\ref{eqn:multiA} zero leads to the eigenvalue equation
for $A_{i}$:
\begin{equation}
[\hat{T}_{i}+V_{i}(\bm{x}_{i},t)+\bm{f}_{i}\cdot\bm{x}_{i}]A_{i}(\bm{x}_{i})=\left(\tilde{E}_{i}+\frac{\bm{P}_{i}^{2}}{2m_{i}}\right)A_{i}(\bm{x}_{i}).
\end{equation}
There are $N$ coupled eigenvalue equations with $N$ position constraints.
Different particles interact with each other under the mean-field
approximation. Now $A_{i}$ is an implicit function of the time $t$
through the explicit dependence on each particle's position:
\begin{equation}
A_{i}(\bm{x}_{i},t)=A_{i}(\bm{x}_{i};\{\bm{X}_{j}(t)\}_{j=1,\cdots,N}).
\end{equation}
Note that the positions of all particles affect the solution of $A_{i}$.

Equation~\ref{eqn:multiExact} transforms to
\begin{align}
\sum_{i}^{N}\frac{\langle\hat{\bm{p}}_{i}\rangle}{m_{i}}\cdot\frac{\dd\langle\hat{\bm{p}}_{i}\rangle}{\dd t} & \approx\langle\frac{\partial V(\bm{x}_{1},\cdots,\bm{x}_{N},t)}{\partial t}\rangle-\frac{\dd}{\dd t}\langle A_{1}\cdots A_{N}|\hat{T}+V(t)|A_{1}\cdots A_{N}\rangle\\
 & =\langle\frac{\partial V(\bm{x}_{1},\cdots,\bm{x}_{N},t)}{\partial t}\rangle-\langle\frac{\partial V(\bm{x}_{1},\cdots,\bm{x}_{N},t)}{\partial t}\rangle\\
 & \quad-\sum_{i}^{N}\frac{\dd\bm{X}_{i}}{\dd t}\cdot[\langle\nabla_{\bm{X}_{i}}(A_{1}\cdots A_{N})|\hat{T}+V(t)|A_{1}\cdots A_{N}\rangle+\nonumber \\
 & \quad\langle A_{1}\cdots A_{N}|\hat{T}+V(t)|\nabla_{\bm{X}_{i}}(A_{1}\cdots A_{N})\rangle]\\
 & =-\sum_{i}^{N}\frac{\langle\hat{\bm{p}}_{i}\rangle}{m_{i}}\cdot\nabla_{\bm{X}_{i}}\langle A_{1}\cdots A_{N}|\hat{T}+V(t)|A_{1}\cdots A_{N}\rangle.\label{eqn:eomMulti}
\end{align}
According to classical mechanics, it is natural to assume that the
change of each $\langle\hat{\bm{p}}\rangle$ should have an opposite
direction to the corresponding energy gradient term ($\nabla_{\bm{X}}\langle A|\hat{H}|A\rangle$),
then the final result is
\begin{equation}
\frac{\dd\langle\hat{\bm{p}}_{i}\rangle}{\dd t}\approx-\nabla_{\bm{X}_{i}}\langle A_{1}\cdots A_{N}|\hat{T}+V(t)|A_{1}\cdots A_{N}\rangle.\label{eqn:eomMulti2}
\end{equation}

\section{Generalization to CNEO-MD}

For model systems above, we did not deal with electronic states, and
the Born-Oppenheimer potential energy surfaces are assumed known beforehand.
For practical molecular systems, we will take the electronic part
into consideration, which will lead to CNEO-MD.

The total Hamiltonian for a molecular system is
\begin{equation}
\hat{H}=\hat{T}_{\text{e}}+\hat{T}_{\text{n}}+\hat{V}_{\text{ee}}+\hat{V}_{\text{nn}}+\hat{V}_{\text{ne}},
\end{equation}
where each term represents electronic kinetic energy, nuclear kinetic
energy, electron-electron interaction energy, nucleus-nucleus interaction
energy and electron-nucleus interaction energy, respectively. Here
for simplicity we ignore the time-dependent external potential but
as shown in sections above, it will not affect the final conclusion.

With the multicomponent Kohn-Sham density functional theory, if nuclei
are assumed distinguishable, the total wave function can be expressed
as
\begin{equation}
|\Psi\rangle=|\phi_{1}\cdots\phi_{N_{\text{e}}}\rangle|\psi_{1}\rangle\cdots|\psi_{N_{\text{n}}}\rangle.
\end{equation}
We can perform similar derivations as in the previous section: starting
with the polar representation for nuclear parts $\psi_{i}$, the nuclear
kinetic energy part $\langle T_{\text{n}}\rangle$ will have an equation
identical to eq~\ref{eqn:kineticMulti}. Without a time-dependent
external field, eq~\ref{eqn:TeqHmVMulti} essentially reduces to
the law of energy conservation:
\begin{equation}
\frac{\dd}{\dd t}E_{\text{tot}}\left[\rho_{\text{e}},\{\psi_{i}\}_{i=1,\cdots,N_{\text{n}}}\right]=0.
\end{equation}
Within the multicomponent Kohn-Sham formalism, if we ignore the electron-nucleus
correlation, the total energy can be divided into the purely electronic
part, the nuclear kinetic energy part, the nucleus-nucleus interaction
part, and the mean-field electron-nucleus interaction part:
\begin{equation}
E_{\text{tot}}\left[\rho_{\text{e}},\{\psi_{i}\}_{i=1,\cdots,N_{\text{n}}}\right]=E_{\text{e}}\left[\rho_{\text{e}}\right]+\langle T_{\text{n}}\rangle+\langle V_{\text{nn}}\rangle+\langle V_{\text{ne}}\left[\rho_{\text{e}}\right]\rangle,
\end{equation}
where the angular brackets represent expectation values taken with
nuclear wave functions. Taking the time derivative gives
\begin{equation}
\frac{\dd}{\dd t}\langle T_{\text{n}}\rangle=-\frac{\dd}{\dd t}\left\{ E_{\text{e}}\left[\rho_{\text{e}}\right]+\langle V_{\text{nn}}\rangle+\langle V_{\text{ne}}\left[\rho_{\text{e}}\right]\rangle\right\} ,
\end{equation}
and
\begin{equation}
\sum_{i}^{N_{\text{n}}}\frac{\langle\hat{\bm{p}}_{i}\rangle}{m_{i}}\cdot\frac{\dd\langle\hat{\bm{p}}_{i}\rangle}{\dd t}=-\frac{\dd}{\dd t}\left\{ E_{\text{e}}\left[\rho_{\text{e}}\right]+\langle V_{\text{nn}}\rangle+\langle V_{\text{ne}}\left[\rho_{\text{e}}\right]\rangle\right\} -\sum_{i}^{N_{\text{n}}}\left\{ \frac{\dd}{\dd t}\langle A_{i}(t)|\hat{T}_{i}|A_{i}(t)\rangle+\frac{\dd}{\dd t}\frac{\sigma_{\bm{p}_{i}}^{2}(t)}{2m_{i}}\right\} .
\end{equation}

Under the adiabatic approximation, the Lagrangian that is used to
perform the constrained minimization is
\begin{align}
\mathcal{L} & =E_{\text{tot}}\left[\rho_{\text{e}},\{\psi_{i}\}_{i=1,\cdots,N_{\text{n}}}\right]-\sum_{ij}\epsilon_{ij}(\langle\phi_{i}|\phi_{j}\rangle-\delta_{ij})\nonumber \\
 & \quad+\sum_{i}^{N_{\text{n}}}\bm{f}_{i}\cdot(\langle\hat{\bm{x}}_{i}\rangle-\bm{X}_{i}(t))-\bm{v}_{i}\cdot(\langle\hat{\bm{p}}_{i}\rangle-\bm{P}_{i}(t))-\tilde{E}_{i}(\langle\psi_{i}|\psi_{i}\rangle-1)\\
 & =E_{\text{e}}\left[\rho_{\text{e}}\right]-\sum_{ij}\epsilon_{ij}(\langle\phi_{i}|\phi_{j}\rangle-\delta_{ij})+\langle A_{1}\cdots A_{N}\left|V_{\text{ne}}\left[\rho_{\text{e}}\right]\right|A_{1}\cdots A_{N}\rangle\nonumber \\
 & \quad+\langle A_{1}\cdots A_{N}\left|\hat{V}_{\text{nn}}\right|A_{1}\cdots A_{N}\rangle+\sum_{i}^{N_{\text{n}}}\langle A_{i}|\hat{T}_{i}|A_{i}\rangle+\frac{1}{2m_{i}}\int\dd\bm{x}_{i}A_{i}^{2}(\bm{x}_{i})[\nabla_{i}S_{i}(\bm{x}_{i})]^{2}\nonumber \\
 & \quad+\bm{f}_{i}\cdot(\langle A_{i}|\hat{\bm{x}}_{i}|A_{i}\rangle-\bm{X}_{i}(t))-\bm{v}_{i}\cdot\left(\int\dd\bm{x}_{i}A_{i}^{2}(\bm{x}_{i})\nabla_{i}S_{i}(\bm{x}_{i})-\bm{P}_{i}(t)\right)-\tilde{E}_{i}(\langle A_{i}|A_{i}\rangle-1).
\end{align}
Similarly, we will have simple solutions to the nuclear phase part
$\nabla_{i}S_{i}(\bm{x}_{i})$ and the nuclear amplitude part needs
to satisfy
\begin{equation}
[\hat{T}_{i}+V_{i}(\bm{x}_{i})+\bm{f}_{i}\cdot\bm{x}_{i}]A_{i}(\bm{x}_{i})=\left(\tilde{E}_{i}+\frac{\bm{P}_{i}^{2}}{2m_{i}}\right)A_{i}(\bm{x}_{i}),\label{eqn:cneoEigen}
\end{equation}
where
\begin{equation}
V_{i}(\bm{x}_{i})=V_{\text{ne}}\left[\rho_{\text{e}}\right](\bm{x}_{i})+V_{\text{nn}}\left[\left\{ \rho_{\text{n},j}\right\} _{j\neq i}\right](\bm{x}_{i}).
\end{equation}

In addition to the nuclear part, the Lagrangian also needs to be stationary
with respect to electronic orbital variations, which gives the electronic
part of multicomponent Kohn-Sham equations:
\begin{equation}
\left\{ \hat{T}_{\text{e}}+U\left[\rho_{\text{e}}\right]+V_{\text{xc}}\left[\rho_{\text{e}}\right]+V_{\text{ne}}\left[\left\{ \rho_{\text{n}}\right\} \right]\right\} \phi_{i}=\epsilon_{i}\phi_{i}.
\end{equation}

The electronic and nuclear Kohn-Sham equations are coupled with each
other and need to be solved self-consistently. This set of constrained
multicomponent Kohn-Sham equations is nothing but the CNEO-DFT equations,
which has been developed in our group.

Similar to the derivations in above sections, the equation of motion
for the observable momentum part can be written as
\begin{align}
\sum_{i}^{N_{\text{n}}}\frac{\langle\hat{\bm{p}}_{i}\rangle}{m_{i}}\cdot\frac{\dd\langle\hat{\bm{p}}_{i}\rangle}{\dd t} & \approx-\frac{\dd}{\dd t}\left\{ E_{\text{e}}\left[\rho_{\text{e}}\right]+\langle A_{1}\cdots A_{N}\left|\hat{T}+\hat{V}_{\text{nn}}+V_{\text{ne}}\left[\rho_{\text{e}}\right]\right|A_{1}\cdots A_{N}\rangle\right\} \\
 & =-\frac{\dd}{\dd t}E_{\text{tot}}\left[\rho_{\text{e}},\left\{ \rho_{\text{n},i}\right\} _{i=1,\cdots,N_{\text{n}}}\right]\\
 & =-\sum_{i}^{N_{\text{n}}}\frac{\dd\bm{X}_{i}}{\dd t}\cdot\nabla_{\bm{X}_{i}}E_{\text{tot}}\left[\rho_{\text{e}},\left\{ \rho_{\text{n},i}\right\} _{i=1,\cdots,N_{\text{n}}}\right]\\
 & =-\sum_{i}^{N_{\text{n}}}\frac{\langle\hat{\bm{p}}_{i}\rangle}{m_{i}}\cdot\nabla_{\bm{X}_{i}}E_{\text{tot}}\left[\rho_{\text{e}},\left\{ \rho_{\text{n},i}\right\} _{i=1,\cdots,N_{\text{n}}}\right].
\end{align}
Following the same arguments on classical mechanics, we have
\begin{equation}
\frac{\dd\langle\hat{\bm{p}}_{i}\rangle}{\dd t}\approx-\nabla_{\bm{X}_{i}}E_{\text{tot}}\left[\rho_{\text{e}},\left\{ \rho_{\text{n},i}\right\} _{i=1,\cdots,N_{\text{n}}}\right].
\end{equation}
This result means that when both electronic and nuclear parts are explicitly
taken into consideration, the whole CMES-MD derivation generalizes
to CNEO-MD.

\section{Numerical validation of the adiabatic wave function propagation}
In this section, we compare the adiabatic wave function propagation with
exact quantum dynamics to test the quality of the adiabatic approximation.
We used the same \ce{^16O^1H} Morse oscillator that was presented in the manuscript for this test.
The initial wave packet is set to
\begin{equation}
    \psi(x,t=0) = A_{\text{gs}}(x) \exp\left(\mathrm{i} \frac{p_0 x}{\hbar}\right),
\end{equation}
where amplitude part $A_{\text{gs}}$ is set to the ground state wave function of the Morse potential,
and key component of the phase part $p_0$ is chosen such that the initial kinetic energy
is equal to $k_{\text{B}}\times 300\,\text{K}$:
\begin{equation}
    \frac{p_0^2}{2m} = k_{\text{B}}\times 300\,\text{K}.
\end{equation}
Here $m$ is the reduced mass of \ce{^16O^1H}.
$p_0$ is chosen to be positive, which means that the initial propagation
stretches the bond.
Because of the nonzero initial observable momentum,
this total wave function is no longer at the ground state, and we simulate
its propagation according to the time-dependent Schr\"odinger equation.

For the CMES-MD simulation, we also start with the ground state wave
function and added the same momentum as in the quantum simulation,
but instead of propagating according to time-dependent Schr\"odinger equation,
the CMES-MD will propagate the position and momentum expectation
values according to Newton's equations but with an underlying quantum wave function picture.

For comparison, we also performed classical MD simulations, it starts with a point charge at the bottom of
the classical potential energy surface (PES),
we also give it the same initial momentum as the other two simulations.

\begin{figure}
    \centering
    \includegraphics[width=\linewidth]{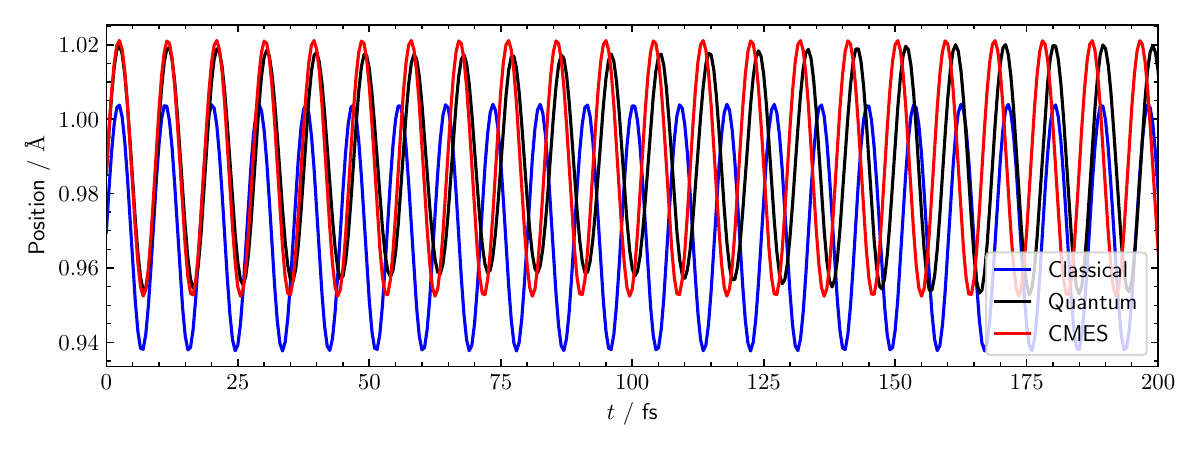}\\
    \includegraphics[width=\linewidth]{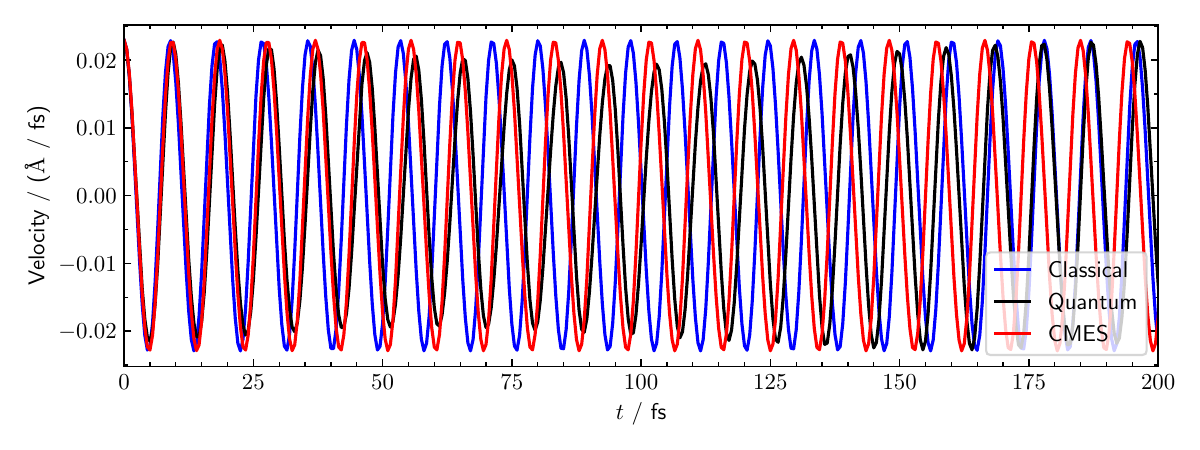}
    \caption{Trajectories from CMES-MD, classical MD and expectation values
    of quantum dynamics for the Morse oscillator.}
  \label{fig:pos_and_vel}
\end{figure}

The (expectation) positions and (expectation) velocities as functions
of time are plotted in Figure~\ref{fig:pos_and_vel}.
Note that CMES and classical PES reach their individual energy minimum at
different bond lengths and therefore initial positions for CMES-MD and classical MD
are different. We can see that CMES-MD performs much better than classical MD
with a very good overlap with the quantum reference.

Next, we directly compare the wave function overlaps. Here two possible definitions for
the overlap are used:
\begin{enumerate}
    \item Total wave function overlap, which is defined as
        \begin{equation}
            I_{\text{tot}}(t) = \left| \int \dd x \psi^*(x, t) A_{\text{CMES}}(x;X(t)) \exp\left(\mathrm{i}\frac{m v(t) x}{\hbar}\right) \right|,
        \end{equation}
        where $\psi(x,t)$ is the wave function according to the quantum dynamics propagation,
        and $A_{\text{CMES}}(x;X(t)) \exp(\mathrm{i}\frac{m v(t) x}{\hbar})$
        is the wave function that is behind the CMES-MD with
        $X(t)$ being the expectation position and $v(t)$ being the expectation velocity.
    \item Amplitude overlap:
        \begin{equation}
            I_{\text{amp}}(t) = \left| \int \dd x A_{\text{QD}}(x, t) A_{\text{CMES}}(x;X(t)) \right|,
        \end{equation}
        where the quantum amplitude is from
        \begin{equation}
            A_{\text{QD}}(x, t) = \sqrt{\left| \psi(x,t)\right|^2}.
        \end{equation}
\end{enumerate}

\begin{figure}
    \centering
    \includegraphics[width=\linewidth]{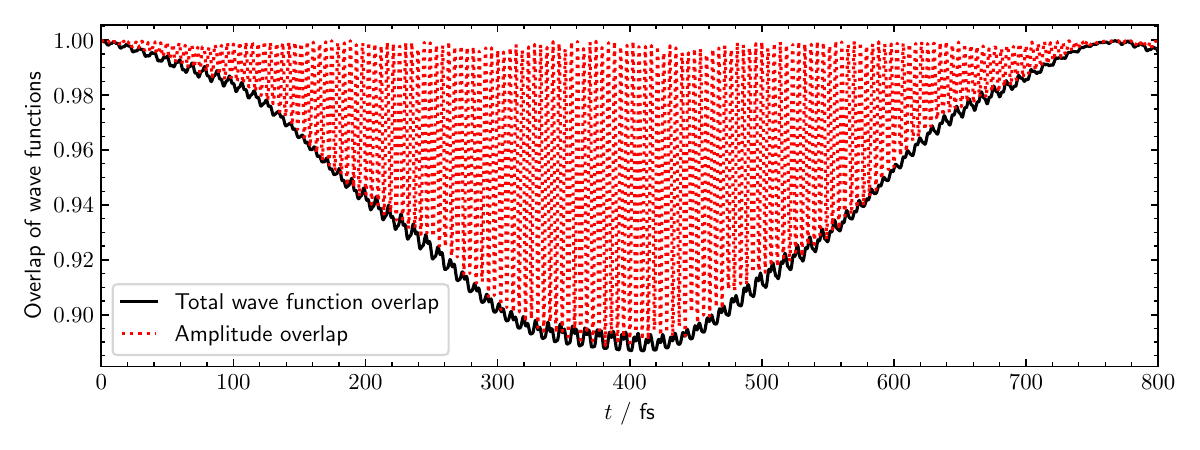}
    \caption{Overlap between the exact time-dependent wave function and CMES
     adiabatic wave functions for the Morse oscillator.}
  \label{fig:ovlp}
\end{figure}

We can see from Figure~\ref{fig:ovlp} that,
the total wave function overlap gradually decreases from 1 to around 0.89
in the first 400\,fs, and then it increases in the next 400\,fs due to a
large period of quantum dynamics. We note that although 0.89 still seems
a big overlap, it already means the physical picture is no longer right
in this system. If we consider an overlap larger than 0.98 to be reasonably
good, then in the short time scale of 100\,fs,
the adiabatic approximation works relatively well, and in a longer time scale,
it cannot match quantum dynamics anymore. The amplitude overlap
roughly describes how good the nuclear density is, and we can see that it
oscillates between approximately 1 (indicating that the wave packet positions
are roughly the same and most errors are in momentum) and the value of
the total wave function overlap (indicating that most errors lie in
the wave packet description in the real space).

\begin{figure}
    \centering
    \includegraphics[width=\linewidth]{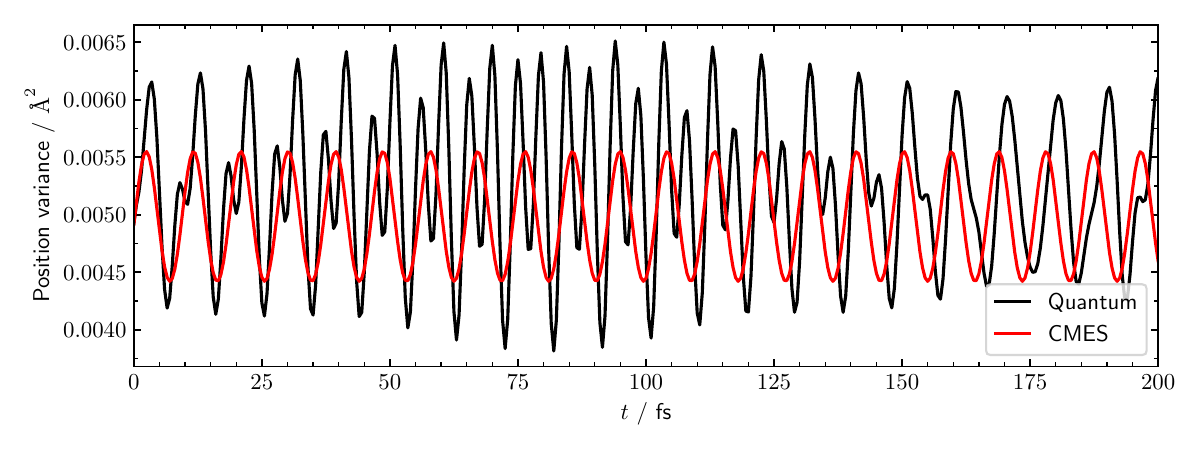}
    \caption{Position variances of the wave functions from CMES-MD and quantum dynamics
    for the Morse oscillator.}
  \label{fig:pos_var}
\end{figure}
Then we investigate a more sensitive property: the variance of position.
It roughly tells how delocalized the wave function is and is heavily
influenced by quantum interferences. Now we can see in Figure~\ref{fig:pos_var}
that
even within the first 100\,fs, the adiabatic CMES-MD is already
much different from quantum dynamics with an oversimplified
dynamics picture and no interference, which is not surprising
considering the classical treatment as well as the adiabatic approximation.

In summary,
CMES-MD can approximate quantum dynamics in a relatively
short time scale, but it can be quite different from the
quantum dynamics in a longer time scale.

\bibliography{manuscript}